\documentclass[
pra, 
aps,
twocolumn,
superscriptaddress,
longbibliography,
floatfix,
notitlepage]{revtex4-2}

\usepackage[utf8x]{inputenc}
\usepackage[english]{babel}
\usepackage[T1]{fontenc}
\usepackage{lmodern}

\usepackage{amsmath,amsfonts,amssymb,amsthm,bm,times,dcolumn,dsfont}

\usepackage{microtype}
\usepackage{braket}
\usepackage{gensymb} 
\usepackage{physics}

\usepackage[colorlinks={true}, citecolor={blue}, filecolor={blue}, linkcolor={blue}, urlcolor={blue}]{hyperref}
\usepackage{graphicx,color}
\usepackage[caption=false]{subfig}
\begin{document}
	
\preprint{APS/123-QED}
	
\title{Steady state entanglement of distant nitrogen-vacancy centers in a coherent thermal magnon bath}%
	
\author{Kamran Ullah}
\email{kullah19@ku.edu.tr}
\affiliation{Department of Physics, Ko\c{c} University, Sar{\i}yer, \.Istanbul, 34450, Turkey}
	
\author{Emre K\"{o}se}
\affiliation{Department of Physics, Ko\c{c} University, Sar{\i}yer, \.Istanbul, 34450, Turkey}
\affiliation{Institut f\"{u}r Theoretische Physik, Eberhard Karls Universit\"{a}t T\"{u}bingen, 72076 T\"{u}bingen, Germany}

\author{Mehmet C. Onba\c{s}l\i}
\email{monbasli@ku.edu.tr}
\affiliation{Department of Electrical and Electronics Engineering,  Ko\c{c} University, Sar{\i}yer, \.Istanbul, 34450, Turkey}
	
\author{\"{O}zg\"{u}r E. M\"{u}stecapl{\i}o\u{g}lu}
\email{omustecap@ku.edu.tr}
\affiliation{Department of Physics, Ko\c{c} University, Sar{\i}yer, \.Istanbul, 34450, Turkey}

\date{\today}
	
\begin{abstract}
We investigate steady-state entanglement (SSE) between two nitrogen-vacancy (NV) centers in distant nanodiamonds on an ultrathin Yttrium Iron Garnet (YIG) strip. We determine the dephasing and dissipative interactions of the qubits with the quanta of spin waves (magnon bath) in the YIG depending on the qubit positions on the strip. We show that
the magnon's dephasing effect can be eliminated, and we can transform the bath into a multimode displaced thermal state using external magnetic fields. 
Entanglement dynamics of the
qubits in such a displaced thermal bath has been analyzed by deriving and solving the master equation. An additional electric field is considered to engineer the magnon dispersion relation at the band edge to control the Markovian character of the open system dynamics. We determine the optimum geometrical parameters of the system of distant qubits and the YIG strip to get SSE. Furthermore, parameter regimes for which the shared displaced magnon bath
can sustain significant SSE against the local dephasing and
decoherence of NV centers to their
nuclear spin environments have been determined. Along with SSE, we investigate the steady-state coherence (SSC) and explain the physical mechanism
of how delayed SSE appears following a rapid generation and sudden death of entanglement
using the interplay of decoherence-free subspace states, system geometry, displacement of the thermal bath, and enhancement of the qubit dissipation near the magnon band edge. A non-monotonic relation between bath coherence and SSE is found, and critical coherence for maximum SSE is determined. Our results illuminate the efficient use of system geometry, band edge in bath spectrum, and reservoir coherence to engineer system-reservoir interactions for robust SSE and SSC. 

\end{abstract}

\maketitle

\section{\label{sec:intro} INTRODUCTION}

Quantum coherence and entanglement are the resources driving the quantum information science and technologies~\cite{Chuang2000}. They are, however, rapidly lost in a system open to environment~\cite{Breuer2007, Zurek2003}. Generating and protecting quantum entanglement, especially steady-state entanglement (SSE), are highly desired. For that aim, interacting two-level systems (qubits) subject to potential or thermal gradients~\cite{Huelga2012,Eisler2014,BohrBrask2015,Hsiang2015,Tavakoli2018,
Tacchino2018,Wang2019,El-Hadidy2019,Tavakoli2020} or time-dependent drives~\cite{Cakir2005,Huelga2007,Li2009,Jin2017} have been examined. Energy-efficient maintenance of nonequilibrium conditions or focusing heat on closely separated qubits are technical challenges that remain to be solved. We follow exactly the opposite route to SSE of two distant qubits in a shared thermal bath~\cite{Braun2002, Benatti2009, Benatti2010, Wolf2011, KongLee2019, Hu2018}. Our approach of bath mediated coupling
between qubits fundamentally differs from proposal that require single-mode system~\cite{Candido_2020}. While shared baths can mediate entanglement between noninteracting qubits, they can suffer from entanglement sudden death (ESD)~\cite{Yu2004}. Adjusting the initial conditions and the bath parameters, a delayed SSE can be revived after ESD~\cite{Yu2004, Orszag2010}. In practice, the qubits could be subject to different local environments in addition to the common bath~\cite{Reiter2013}. We specifically investigate the interplay of an external field engineered shared bath and the geometry of the bath-qubits system to beat ESD for retrieving delayed SSE effect in the presence of other local environments.

Our system consists of an ultrathin Yttrium Iron Garnet (YIG) nanostrip~\cite{Ding2020,Zhang2014, Hauser2016,Collet2017,Savchenko2019,Talalaevskij2017,Jungfleisch2015,Klingler2015,HoaiHuong2020} and two distant (non-interacting) nanodiamonds hosting nitrogen-vacancy (NV) center defect qubits as illustrated in Fig.~\ref{fig:fig1-ModelSystem}. Such a system of NV centers and YIG strip waveguide is shown to be promising for long distance scalable entanglement generation in transient regime~\cite{Fukami2021}. Qubits couple to spin waves in the garnet. Weak excitations of spins 
about the $z$-axis are described as bosonic quasiparticles, magnons~\cite{Bloch1930,Holstein1940,Dyson1956}. In broader context, hybrid systems of qubits in magnon baths play a central role in the field of magnonics~\cite{Lachance-Quirion2019,Bertelli2020,Lee-Wong2020,Gonzalez-Ballestero2020}.
Our geometrical parameters are the qubit positions and the dimensions of the strip.
We use an effective linear spin chain in the $x$-direction to represent the 
magnetic strip to calculate bath-qubits couplings. Two static magnetic fields, $\boldsymbol{B_0},\boldsymbol{B_1}$ are assumed to be applied perpendicular to the strip, in the $z$ and $-y$ directions, respectively.
We take the magnons as a thermal bath but it is ``displaced" and injected quantum coherence by the $\boldsymbol{B_1}$ field.
Accordingly, our model describes two spin qubits immersed in a quasi-one-dimensional displaced thermal bath of magnons. Very recently, thermal control of brodband magnons in YIG crystals has been proposed~\cite{Fung_2021}. Further control on the magnon dispersion relation is introduced by an electric field transverse to the YIG axis~\cite{Mills2008,Liu2011,Liu2012,Krivoruchko2018,Krivoruchko2018b,Savchenko2019,Candido_2020,Fung_2021}. 

If two non-interacting qubits are in a common bath of magnons, and if one qubit is excited while the other one is in its ground state initially, then the excitation (energy) is exchanged between the qubits by the magnons. Hence, magnon mediated interaction between the qubits is the essential physics that can yield SSE. External fields, optimization of system geometry, and bath engineering however are required to realize SSE in real systems where additional
local baths to qubits can be present. To examine the open system dynamics, we derive the master equation of the open qubit system by carefully discussing the Born, Markov, and secular approximations~\cite{Breuer2007}, taking
into account the geometry dependence of interaction coefficients between the magnons and the qubits.
We find the structure of our master equation is similar to the squeezed thermal bath master equation for a driven system used
for ESD and delayed SSE generation schemes~\cite{Orszag2010}, when the qubits are placed away from
the ends of the strip. In contrast to weak squeezing that may arise from nonlinear higher order interactions, the effective squeezing in the displaced bath can be large and controlled by the external static field $\boldsymbol{B_1}$. Furthermore, the dissipation rates to the public  bath is enhanced at the band edge of the magnonic crystal, which allows for SSE even in the presence private baths of the qubits, similar to the enhancement of radiative decay rates in photonic crystals~\cite{Vats1998,Roy2010,Wang2011,Wang2012,Yang2013,Wang2014,Woldeyohannes2015,Li2015,Wu2016,Shen2019}. The coherence injected by $\boldsymbol{B_1}$ into the thermal bath, contributes to both local and nonlocal dissipators; besides, it generates an effective drive term
on the qubits. Hence, a non-monotonic effect of coherence on SSE is predicted due to the competing roles it plays in the dynamical processes. We determine the critical coherence for maximum SSE. Moreover, we point out
a subtle interplay of the system geometry with the special qubit states spanning a decoherence-free subspace (DFS)~\cite{Lidar2012} for the system-bath interactions to get SSE. 

In addition to SSE generation and protection, we discuss the steady-state coherence (SSC) structure of the two-qubit states
explicitly. We find that significant coherence is generated robustly along
with the entanglement, even in parameter regimes where entanglement is weak
or does not exist. The generated coherences in the qubit pair are versatile, 
significant beyond typical quantum information applications, such as
quantum information and heat engines~\cite{Dag2016,Tuncer2020,Latune2019,Latune2019b,Latune2020}.
Our scheme can be relatively easier to implement in comparison to schemes requiring precise timing of external pulses as it does not require time-dependent drives; besides in comparison to typical bath induced entanglement generation using private baths, common bath is not subject to the problem of focusing thermal noise onto qubits locally. In addition, 
our scheme can be scalable by placing more qubits on the YIG strip straightforwardly for multipartite SSE and SSC generation and protection for diverse quantum technology applications for quantum metrology, simulations, or computations.

The rest of the paper is organized as follows.
In Section~\ref{sec:modelSystem}, we describe our model system consisting 
of a YIG nanostrip and a pair of NV-center qubits, the interactions
between the qubits, and the displaced magnon bath in three subsections.
In Section~\ref{sec:results}, first two subsections
present the justification of system parameters and the resonance condition between the NV centers and the magnetostatic magnon mode. Third subsection presents the spatial profile of coherence function of the bath modes, and the derivation
of the master equation for the open system of qubits is given in the fourth subsection. Fifth subsection presents the SSE results in three parts. First is the case of SSE generation and protection when decoherence channels of the qubits to their local nuclear spin environments are neglected. Second, the local decoherence channels of the qubits are included to present how the ESD is compensated by the squeezing effect of common displaced environment to achieve SSE. Third, the role of DFS for SSE with and without coherence in the magnon bath is discussed. We conclude in Section~\ref{sec:conc}.

\section{Dynamics of our model system: A pair of NV centers on a YIG nanostrip}
\label{sec:modelSystem}

\subsection{YIG nanostrip and displaced thermal magnon bath}
\label{sec:model-magnonBath}

\begin{figure}[t!]
\centering
\includegraphics[width=\linewidth]{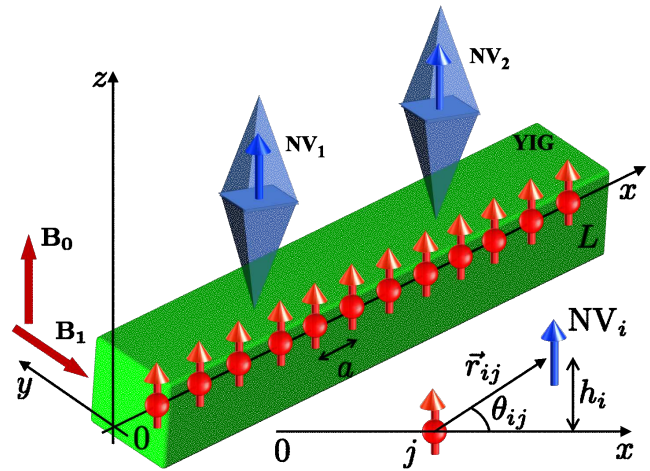}
\caption{\label{fig:fig1-ModelSystem} (Color online) Schematic view of a pair of 
NV center spins in nanodiamonds (blue arrows inside octahedrons) on an linear spin chain (red arrows with spheres) of length $L$ along the $x$-axis, effectively modeling a YIG nanostrip. Static magnetic fields $\bm{B_0}$ and $\bm{B_1}$ are applied in the $z$ and $-y$ directions, respectively. It is assumed that $B_1$ is focused on YIG and negligible on the NV centers. An electric field (not shown) can be further considered transverse to the $x$ axis to control the effective thickness of the YIG strip. The distance between nearest-neighbor spins is denoted by $a$. NV center, labeled by $i=1,2$ is at a height $h_i$ from the chain, making an angle $\theta_{ij}$ with the vector $\bf{r_{ij}}$ connecting to the $j$-th spin of the linear chain.}
\end{figure}

We consider a YIG, Y$_3$Fe$_2$(FeO$_4$)$_3$, nanostrip that hosts our magnon bath, in external magnetic and electric fields, as illustrated in Fig.~\ref{fig:fig1-ModelSystem}. Microfabricated ultrathin YIG films~\cite{Hauser2016}, YIG strips, and waveguides~\cite{Collet2017,Savchenko2019} are experimentally available.
YIG crystals can be grown with high purity, and they can maintain spin waves with low damping and acoustic dissipation rates. Magnons are the quanta of such spin waves, described by an Hamiltonian
\begin{eqnarray}\label{eq:H_magnon_0}
\hat H_{\text{mag},0}=\hbar\sum_{k=-\infty}^{\infty} 
\omega_k \hat m^\dagger_k \hat m_k,
\end{eqnarray}
where $\hat m_k$ ($\hat m^\dagger_k$) is the annihilation (creation) operator of a magnon quasiparticle with wavenumber $k$ and frequency $\omega_k$ (A short introduction to magnons is presented in Appendix \ref{sec:appendix-magnons}).

Though YIG is a ferrimagnet with a complex lattice structure, it has a well-separated ferromagnetic lowest band, described by Heisenberg  exchange interactions of effective spins $\hat{\boldsymbol{S}}_i=(\hat{S}_i^x,\hat{S}_i^y,\hat{S}_i^y)$ at the sites $i$ on an effective simple cubic lattice with the lattice constant $a=12.376$ \AA.
Saturation magnetization of the bulk YIG crystal is $\mu_0 M_s= 175$ mT, which
gives the magnitude of the effective spin  $s$ for a simple cubic unit cell block as $s=14.2$, from the definition of the magnetization $M_s=\mu/a^3\sim 140$ kA/m. Here, $\mu=g\mu_Bs$, $\mu_B$ is
the Bohr magneton, and the effective g-factor is $g=2$. 
The value of $s$ changes slightly with the width of the YIG strip, for example, it becomes $s=10.21$ for a $20$ nm width YIG strip~\cite{Collet2017,Savchenko2019} for which $\mu_0M_s\sim 100$ mT~\cite{Talalaevskij2017,Jungfleisch2015}.

The ferromagnetic exchange interaction, characterized with positive strength 
is short-ranged and only couples the nearest-neighbor sites.
It is calculated by using the measured exchange stiffness constant $A=3.7\pm0.4$ pJ/m~\cite{Klingler2015} and the
relation of the $\rho_s$ to the magnon dispersion relation via 
$J= A a/s^2$. We find $J/2\pi=33.42$ GHz. Spin stiffness varies weakly (within $10\%$) with the temperature, unless close to the Curie temperature $T_\mathrm{C}$, where it sharply drops to zero~\cite{HoaiHuong2020}.
Remarkably, one could consider doping YIG crystal to get significant enhancement to
the coupling coefficient even close to the $T_\mathrm{C}$~\cite{HoaiHuong2020}. 

The large magnitude of effective spin $s\sim 14.2$ associated with the effective cubic unit cell description of YIG crystal allows us to employ classical dispersion relation together with our microscopic chain model~\cite{Liu2011}. In the case of a finite width quasi-one-dimensional YIG strip, subject to transverse magnetic and electric fields, the dispersion relation is given by~\cite{Kalinikos1986,Jorzick2001,Zhang2016,Dieterle2019}
\begin{eqnarray}\label{eq:magnonDispersionFiniteSize}
\omega_n(k)=\sqrt{\omega_{an}(k)\omega_{bn}(k)}-v_Ek
\end{eqnarray}
where we introduced short hand notations,
\begin{eqnarray}
\omega_{an}(k)&=&\omega_0+2Jsa^2k_n^2,\\
\omega_{bn}(k)&=&\omega_0+2Jsa^2k_n^2+\omega_M(1-\frac{1-\text{e}^{-k_nL_z}}{k_nL_z}).
\end{eqnarray}
Here,  $\gamma_0=g\mu_B/\hbar$ is the gyromagnetic ratio (in units of rad/Ts), and $\omega_0:=\gamma_0B_0$, $\omega_M:=\gamma_0\mu_0 M_s$, 
and $v_E:=\omega_M L_E$, with
\begin{eqnarray}
L_E:= \frac{4\gamma_0 A |e| E}{\omega_M M_s E_{\mathrm{SO}}}.
\end{eqnarray}
We denote $k_n^2:=k^2+(n_y\pi/L_y)^2$, with $n_y=0,1,2,...$ and $k\equiv k_x$. $e$ stands for the electron charge. $E_{\mathrm{SO}}\sim 19$ eV $\sim 3.044$ aJ is an energy scale related to the inverse of the Dzyaloshinskii-Moriya (DM) interaction coefficient, reflecting the microscopic spin-orbit coupling effect~\cite{Mills2008,Liu2011,Liu2012,Krivoruchko2018,Krivoruchko2018b,Savchenko2019}.
We assume electric field is transverse to the YIG strip axis ($x$) and NV centers
are shielded from its influence. Its main purpose is to control the group
velocity for the magnetostatic (long wavelength) modes, which in return affects the magnon bath dissipation rates through the magnon DOS. 

In what follows, we drop the mode index $n=0$. From $\omega(k)$ we can calculate
the DOS, which becomes 
\begin{eqnarray}\label{eq:dosStripE}
D(\omega_0)\equiv D_0=\frac{8L_x}{\omega_M(L_z-L_E)},
\end{eqnarray}
at $k=0$. Denominator of Eq.~(\ref{eq:dosStripE}) can be interpreted as an effective geometrical role played by the electric field. $L_E$ allows us to effectively make the YIG strip thinner for the purpose of controlling the DOS at the magnetostatic modes.
Remarkably, when $E=0$, $D_0\sim 10^{-8}$ s; using high electric field 
$E\sim 0.1$ V/nm and high precision tuning between $L_z$ and $L_e$ we can get
$D_0\sim 0.25$ s. Another subtle point is that
the dispersion relation is no longer an even function of $k$, and the summations over $k$ should be from $-\infty$ to $+\infty$ and hence the directional degeneracy factor in the DOS is not employed.

We further consider a static uniform field $\boldsymbol{B_1}$ is applied to the YIG nanostrip in the $y$-axis, whose purpose is to make SSE more robust against additional decoherence channels.  
An additional Zeeman term for $\boldsymbol{B_1}$  
in the $-y$ direction, in terms of magnon operators, is added to magnon
Hamiltonian~(\ref{eq:H_magnon_0}),
\begin{eqnarray}\label{eq:Hdis}
\hat H_{\text{mag},1}
=i\hbar \sum_{k=-\infty}^{\infty}({\cal E}_k\hat{m}_k^\dag-{\cal E}_k^\ast\hat{m}_k),
\end{eqnarray}
where
\begin{eqnarray}\label{eq:coh_k}
{\cal E}_k=\gamma_0\sqrt{\frac{s}{2N}}\sum_{j=-N/2}^{N/2} B_{1j}\text{e}^{-ik x_j}.
\end{eqnarray}
Here, $B_{1j}$ is the magnitude of the magnetic field on the spin
site $x_j$. We consider only static fields, and do not aim to excite a particular spin wave mode. Our approach may have some practical advantages for implementations as we do not require precise 
timing of time-dependent drive fields in our theory
and we get SSE and SSC through natural relaxation of the open system in contrast to external dynamical control schemes.

For simplicity, we only consider a single linear spin chain to estimate
the injected coherence (displacement) to the magnons. Spin locations are given by
\begin{eqnarray}
x_j=[j-\text{sign}(j)\frac{1}{2}]a,
\end{eqnarray}
with the sign function, $\text{sign}(x)=+1,0,-1$ for $x>0,x=0,x<0$, respectively.

We treat the magnon subsytem with a wide and continuous spectrum, except the gap
at $k=0$, as a large bath to the NV centers. Its initial state can be determined solely by its own total Hamiltonian
\begin{eqnarray}\label{eq:Hmag}
\hat H_{\text{mag}}=
\hbar\sum_{k=-\infty}^{\infty} (\omega_k \hat m^\dagger_k \hat m_k
+i({\cal E}_k\hat{m}_k^\dag-{\cal E}_k^\ast\hat{m}_k)),
\end{eqnarray}
and the thermal environment, which we do not specify its coupling to the magnons 
except assuming that it would bring the magnons to a thermal equilibrium, if there
would be no coherence at a temperature $T$. In the case of coherence, 
we first diagonalize the magnon Hamiltonian by using the multimode 
Glauber displacement operator with the coherence parameter $\epsilon_k$~\cite{Glauber1963}
\begin{eqnarray}
\hat{D}(\epsilon_k)=\exp(\epsilon_k \hat{m}_k^\dag-\epsilon_k^\ast \hat{m}_k).
\end{eqnarray}
For $\epsilon_k=-i{\cal E}_k/\omega_k$ we find
\begin{eqnarray}\label{eq:HmagDis}
\hat H_{\text{mag}}=
\hbar\sum_{k=-\infty}^{\infty}\omega_k \hat m^{\prime\dagger}_k \hat m_k^\prime,
\end{eqnarray}
where $\hat{m}^{\prime}_k=\hat{m}_k-\epsilon_k$ and a constant of $|\epsilon_k|^2$ is dropped. In what follows, we suppress the prime superscripts for brevity.

The magnetic field amplitude $B_1$ must be less than than the maximum field that
would saturate the magnetic material along the $y$-axis. Saturation field can be controlled and
can be high ($\sim 0.5$ T) in YIG materials with perpendicular magnetic anisotropy (PMA), which can be physically implemented by substrate strain or replacing yttrium with other rare earth ions~\cite{Wang2014a,Fu2017,Li2019,Guo2019,MokarianZanjani2020,Ding2020}.
Maximum value of $B_1$ limits how much coherence can be injected to the magnons.
For example, in a YIG nanostrip with $N\sim 10^3$ sites along the
long axis, the range of coherence of the magnetostatic mode ($k=0$) becomes $|\epsilon_0|<\sim 1$, taking $B_0=51.16$ mT. The value of $B_0$ is fixed by the resonance condition in Sec.~\ref{sec:resonance}.

If we assume that the spin chain is in contact with a thermal environment then
the magnon reservoir is described as a coherent (displaced) thermal bath for the
NV centers, with the correlations
\begin{eqnarray}\label{eq:DisplacedThermalBath-mmdag}
\expval{\hat{m}_k}&=&-\epsilon_k,
\label{eq:DisplacedThermalBath-m}\\
\expval{\hat{m}_k \hat{m}_q}&=&\epsilon_k\epsilon_q,
\label{eq:DisplacedThermalBath-mm}\\
\expval{\hat{m}_k^\dag \hat{m}_q}&=&\delta_{kq}\bar{n}_k + 
\epsilon_k^\ast\epsilon_q, 
	\label{eq:DisplacedThermalBath-mdagm}\\
\expval{\hat{m}_k \hat{m}_q^\dag}&=&\delta_{kq}(\bar{n}_k + 1) + 
\epsilon_k\epsilon_q^\ast,	
\end{eqnarray}
where the thermal contribution to the mean number of magnons is given by
the Bose-Einstein distribution function
\begin{eqnarray}
\bar{n}_k(T)=\frac{1}{\exp(\hbar\omega_k/k_BT)-1},
\end{eqnarray}
with $k_B$ being the Boltzmann constant. 

\subsection{Diamond NV center qubits}
\label{sec:NVcenterQubits}

Hamiltonian of the NV center qubits is derived in 
Appendix~\ref{sec:appendix-NVqubits} and it is given by
\begin{eqnarray}\label{eq:HnvQubit}
\hat H_{\mathrm{NV}}=\hbar\frac{\omega_{\mathrm{NV}}}{2}\sum_{i=1,2}
\hat\sigma_i^z,
\end{eqnarray}
where $\hbar\omega_{\mathrm{NV}}:=\hbar(D-\gamma_\mathrm{NV}B_0)$ and $\hat{\sigma}^z_i:=|-1\rangle_i\langle -1|-|0\rangle_i\langle0|$. In the subsequent discussions we use 
$\hat{\sigma}_i^+=
\ket{-1}_i\bra{0}_i$ and $\hat{\sigma}_i^-=\ket{0}_i\bra{-1}_i$. For simple analytical expressions, we assumed $\boldsymbol{B_1}$ is negligible on the NV centers. This is not a prerequisite for any experimental implementation of our proposal, and $\boldsymbol{B_1}$ does not have to be focused on the YIG only. For a practical realization that cannot have negligible $\boldsymbol{B_1}$ on the qubits, one can simply diagonalize the NV center Hamiltonian when $\boldsymbol{B_1}$ is present to find the corresponding qubit transition frequency $\omega_{\mathrm{NV}}$. The essential contribution of $\omega_{\mathrm{NV}}$ in the rest of the theory is to determine the
magnitude of $\boldsymbol{B_0}$ to satisfy the magnon-qubit resonance, which would depend on the given $\boldsymbol{B_1}$ magnitude. As $\boldsymbol{B_1}$ determines
the injected coherence, one would have different resonance fields
for different coherences. Other than this minor technical change, the open system dynamics and the essential physics of SSE and SSC generation remains the same and hence we continue with neglecting  $\boldsymbol{B_1}$ on the NV centers.

\subsection{NV center - magnon interactions}
\label{sec:model-spinChainMagnons}

Let us consider a YIG strip of thickness $L_z$, width $L_y$, and length $L\equiv L_x$, with conditions
$L_z\ll L_y \ll L_x$. For simplicity, we consider an effective one-dimensional spin chain as a close representation of the ultrathin YIG strip to calculate its coupling to the NV centers. 
Our effective spin chain
corresponds to a linear lattice of cubic unit cells, and hence, it is associated a width of $a\sim 1$ nm. An ultrathin nanostrip could have a few nm thickness and width of $L_y\sim 10$ nm so that
a more rigorous calculation would need to consider several spin chains symmetrically placed
next to the central one. We expect the overall effect of neigboring chains could
yield a collective enhancement of the interaction coefficients we estimate here.
We limit ourselves to an underestimation of the interaction coefficients for the sake of avoiding additional complexity in our theoretical treatment. 

Interaction between an NV-center qubit represented by a spin $\hat{\boldsymbol{\sigma}}_i$ with $i=1,2$ and 
a spin $\hat{\boldsymbol{S}}_j$ at a cite $j$ in the effective linear chain representing the YIG nanostrip is given by the magnetic dipolar coupling 
\begin{eqnarray}\label{eq:Hdipolar}
H_{\mathrm{int}}^{(ij)}=\hbar d_{ij} \left[
 \bm{\sigma}\cdot\bm{S}_j-3(\bm{\sigma}_i\cdot\bm{e}_{ij})
 (\bm{S}_j\cdot\bm{e}_{ij}) \right],
\end{eqnarray}
where, $\boldsymbol{e}_{ij}=\boldsymbol{r}_{ij}/{r}_{ij}$ is the unit 
vector in the direction of the distance vector
$\boldsymbol{r}_{ij}={r}_{ij}(\cos{\theta}_{ij},\sin{\theta}_{ij})$ from the chain 
site $j$ to the NV center in the $xz$-plane, as shown in Fig.~\ref{fig:fig1-ModelSystem}. The coefficient 
$d_{ij}:=\hbar\mu_0 \gamma_{\mathrm{NV}}\gamma_0/8\pi r_{ij}^3$ is the frequency of dipolar coupling. The angle $\theta_{ij}$ is between
the $\boldsymbol{r}_{ij}$ and the $x$-axis so that $r_{ij}=z_i/\sin\theta_{ij}$ with $z_i$ is the height of the ith NV center from the spin chain. For simplicity
we take $z_1=z_2\equiv z_{\mathrm{NV}}$ and write $d_{ij}=d\sin^3\theta_{ij}$ with $d=\hbar\mu_0 \gamma_{\mathrm{NV}}\gamma_0/8\pi z_{\mathrm{NV}}^3$.

We can find the magnon representation of
Eq.~(\ref{eq:Hdipolar}) by writing it in terms of the ladder operators 
$\hat{S}_{j}^{\pm} = \hat{S}_{j}^x \pm i\hat{S}_{j}^y$,
$\hat{\sigma}_i^{\pm}=(\hat{\sigma}_i^{x} \pm i\hat{\sigma}_i^{y})/2$, and using Eqs.~(\ref{eq:HolsteinPrimakoff1})-(\ref{eq:HolsteinPrimakoff2}). In addition to the bilinear $\hat{\sigma}_i^{\pm,z}\hat m_j$ and $\hat{\sigma}_i^{\pm,z}\hat m_j^\dagger$ terms, the dipolar interaction gives rise to terms that only depends on NV center operators. Coefficient of $\sigma_i^z$ shifts the NV center frequencies to 
\begin{eqnarray}\label{eq:omega_i}
\omega_{i}=\omega_\text{NV}-\sqrt{2s}\beta_i,
\end{eqnarray}
where
\begin{eqnarray}
\beta_i:=\sum_{j=-N/2}^{N/2}2B_{ij},
\end{eqnarray}
with 
\begin{eqnarray}
B_{ij}:=-d\frac{\sqrt{2s}}{2}\sin^3{\theta_{ij}}(3\cos^2{\theta_{ij}}-2).
\end{eqnarray}
Coefficients of $\hat{\sigma}_i^{\pm}$ describe
NV center transitions driven by classical spin waves.
Combination of these dipolar interaction terms
with Eq.~(\ref{eq:HnvQubit}) yields a
Hamiltonian
\begin{eqnarray}\label{eq:HnvPrime}
\hat{H}_\text{NV}^\prime=\hbar
\sum_{i=1,2}\left[\frac{\omega_{i}}{2}\hat\sigma_i^z
-
\sqrt{2s}\alpha_{i}
(\hat{\sigma}_i^{+}+\hat{\sigma}_i^{-})\right],
\end{eqnarray}
where
\begin{eqnarray}
\alpha_{i}:=\sum_{j=-N/2}^{N/2}A_{ij},
\end{eqnarray}
with
\begin{eqnarray}
A_{ij}:=
-d\frac{3\sqrt{2s}}{4}\sin^3{\theta_{ij}}
\sin2\theta_{ij}.
\end{eqnarray}
We introduced $\alpha_i, \beta_i, A_{ij}$,and $B_{ij}$ notations for brevity, as they will appear in other terms in the total Hamiltonian, too.

The rest of terms in Eq.~(\ref{eq:Hdipolar}) can be 
grouped into three different types of magnon-qubit
interactions expressed as
\begin{eqnarray}
\hat{H}_\text{deph}&=&\hbar\sum_{ij}A_{ij}
\hat{\sigma}_i^z
(\hat{m}_j^\dagger+\hat{m}_j),
\label{eq:HdephSiteRep}\\
\hat{H}_\text{crt}&=&\hbar\sum_{ij}B_{ij}
(\hat{\sigma}_i^-\hat{m}_j+\text{H.c.}),
\label{eq:HcrtSiteRep}\\
\hat{H}_\text{rt}&=&\hbar\sum_{ij}C_{ij}
(\hat{\sigma}_i^-\hat{m}_j^\dagger+\text{H.c.}).
\label{eq:HrtSiteRep}
\end{eqnarray}
The Hamiltonian $\hat{H}_\mathrm{deph}$ is responsible for the NV qubit dephasing. The counter rotating 
terms (crt) and
rotating terms (rt) are collected into the $\hat{H}_\mathrm{crt}$ and $\hat{H}_\mathrm{rt}$, respectively. The coefficient $C_{ij}$ is defined to be
\begin{eqnarray}
C_{ij}=-d\frac{3\sqrt{2s}}{2}\sin^3{\theta_{ij}}
\cos^2{\theta_{ij}}.
\end{eqnarray}

We will rotate the NV qubit basis 
$\ket{-1}_i,\ket{0}_i$ to a new one $\ket{-}_i,\ket{+}_i$
\begin{eqnarray}
\ket{+}_i&=&\cos{\phi_i}\ket{-1}_i+\sin{\phi_i}\ket{0}_i,\\
\ket{-}_i&=&-\sin{\phi_i}\ket{-1}_i+\cos{\phi_i}\ket{0}_i,
\end{eqnarray}
to diagonalize the
Hamiltonian in Eq.~(\ref{eq:HnvPrime}). The basis rotation translates
into the $2\phi_i$ rotation about the $y$-axis of the NV qubit spins
so that we have
\begin{eqnarray}
\hat{\sigma}^z_i&\rightarrow
&\hat{\sigma}^z_i\cos{2\phi_i}-\hat{\sigma}^x_i\sin{2\phi_i}
,\\
\hat{\sigma}^x_i&\rightarrow
&\hat{\sigma}^z_i\sin{2\phi_i}+\hat{\sigma}^x_i\cos{2\phi_i},\\
\hat{\sigma}^+_i&\rightarrow
&\frac{\hat{\sigma}^z_i}{2}\sin{2\phi_i}+\hat{\sigma}^+_i\cos^2{\phi_i}-\hat{\sigma}^-_i\sin^2{\phi_i}.
\end{eqnarray} 
Here the spin operators on the right hand side are in the $\ket{\pm}$ basis
such that 
$\sigma_i^z\equiv\ket{+}_i\bra{+}-\ket{-}_i\bra{-}$ and 
$\hat{\sigma}_i^{\pm}=\ket{\pm}_i\bra{\mp}$. 

We find that at an angle of rotation determined by the condition
\begin{eqnarray}
\tan2\phi_i=\frac{2\sqrt{2s}\alpha_i}
{\sqrt{2s}\beta_i-\omega_\text{NV}},
\end{eqnarray}
Eq.~(\ref{eq:HnvPrime}) becomes diagonal in the $\ket{\pm}$ basis, 
\begin{eqnarray}\label{eq:HnvNewBasis}
\hat{H}_\text{NV}=
\hbar\sum_{i=1,2}\frac{\Omega_i}{2}\hat\sigma_i^z,
\end{eqnarray}
where we dropped the prime such that $\hat{H}_\text{NV}^\prime\equiv\hat{H}_\text{NV}$.
The new qubit transition frequency is 
\begin{eqnarray}\label{eq:dressedNVfreq}
\Omega_i:=(\omega_i^2+8s\alpha_i^2)^{1/2}.
\end{eqnarray}

In terms of the new NV qubit spin operators, the interaction terms
can be found similarly. We get exactly the same form of 
interaction Hamiltonians as in Eqs.~(\ref{eq:HdephSiteRep})-(\ref{eq:HrtSiteRep}), but the
interaction coeffients 
$A_{ij},B_{ij},C_{ij}$ are replaced by
$\xi_{ij},\zeta_{ij},\eta_{ij}$, respectively, where
\begin{eqnarray}
\xi_{ij}&=&A_{ij}\cos2\phi_{i}
+\frac{1}{2}(B_{ij}+C_{ij})\sin2\phi_{i},\\
\zeta_{ij}&=&-A_{ij}\sin2\phi_{i}
+\frac{B_{ij}+C_{ij}}{2}\cos2\phi_{i}\nonumber\\
&+&\frac{B_{ij}-C_{ij}}{2},\\
\eta_{ij}&=&-A_{ij}\sin2\phi_{i}
+\frac{B_{ij}+C_{ij}}{2}\cos2\phi_{i}\nonumber\\
&-&\frac{B_{ij}-C_{ij}}{2}.
\end{eqnarray}

To express the Hamiltonians in $k$-space we use
\begin{eqnarray}\label{eq:superposition}
f_k^{(i)}&=&\frac{1}{\sqrt{N}}\sum_{j=-N/2}^{N/2} 
f_{ij}\text{e}^{-ik x_j},
\end{eqnarray}
where $f\in\{\xi_{ij},\zeta_{ij},\eta_{ij}\}$. Accordingly,  Eqs.~(\ref{eq:HdephSiteRep})-(\ref{eq:HrtSiteRep})
become
\begin{eqnarray}
\hat{H}_\text{deph}&=&
\hbar\sum_{ik}\xi_{k}^{(i)}\hat{\sigma}_i^z\hat{m}_k
+\text{H.c.},
\label{eq:Hdeph_kRep}\\
\hat{H}_\text{crt}&=&\hbar\sum_{ik}\zeta_{k}^{(i)}
\hat{\sigma}_i^-\hat{m}_k
+\text{H.c.},
\label{eq:Hcrt_kRep}\\
\hat{H}_\text{rt}&=&\hbar\sum_{ik}\eta_{k}^{(i)}
\hat{\sigma}_i^+\hat{m}_k
+\text{H.c.}.
\label{eq:Hrt_kRep}
\end{eqnarray}

Together with the Eq.~(\ref{eq:HnvNewBasis}), and  
Eq.~(\ref{eq:coh_k}), Eqs.~(\ref{eq:Hdeph_kRep})-(\ref{eq:Hrt_kRep}) complete the total Hamiltonian $\hat H$ of the overall system expressed in $k$-space. Hamiltonian
can be written in $\omega$ space as well by using the magnon
DOS. The interaction coefficients are highly sensitive to the geometry of the
setup. Remarkable differences emerge between the central and closer to edges placements of the NV centers on the chain. The decoherence and dephasing rates of the NV qubits to the common magnon bath are determined by the
interaction coefficients. Hence, the geometric dependence of the interaction coefficients is translated to the open system dynamics of the NV center qubits. To see the explicit relation of geometry and open system dynamics, our next aim is to develop the master equation of the system.

\section{RESULTS and DISCUSSION}\label{sec:results}
Initially, the qubit system is assumed to be prepared in a state where only one of the qubits is excited, $\rho(0)=\ket{+-}\bra{+-}$. This ensures bath mediated energy exchange could be established between the qubits through the nonlocal dissipator of the public (common) bath. We propagate the qubit state by 
solving the master equation and then 
determine their entanglement dynamics by calculating the bipartite
concurrence~\cite{Wootters1998}
\begin{eqnarray}
C=\text{max}\{0,\sqrt{\lambda_1}-\sqrt{\lambda_2}-\sqrt{\lambda_3}-\sqrt{\lambda_4}\}.
\end{eqnarray}
Here, the eigenvalues $\lambda_i$ with $i=1..4$ of the time-reversed matrix $R=\rho\tilde{\rho}$ 
are in the descending order, where 
$\tilde{\rho}=(\sigma^y\otimes\sigma^y)\rho^\ast(\sigma^y\otimes\sigma^y)$ is the spin flipped density matrix.
We use the standard basis $\{\ket{1}\equiv\ket{11},\ket{2}\equiv\ket{10},\ket{3}\equiv\ket{01},\ket{4}\equiv\ket{00}\}$ with $\ket{+}\equiv\ket{1}$ and $\ket{-}\equiv\ket{0}$. 

In addition, dynamical behavior of the entanglement is compared to the coherence, which is quantified by the $l_1$ norm coherence~\cite{Baumgratz2014}
\begin{eqnarray}
C_{l_1}(\rho) \equiv C_1:=\sum_{\substack{i,j\\ i\neq j}}|\rho_{ij}|.
\end{eqnarray}

We first discuss the rest of the physical parameters required for our simulations and derivation of the master equation, then present our
results in the following subsections.

\subsection{Physical parameters}\label{sec:physParameters}

NV center qubits in diamond hosts can be found at heights of $~5-100$ nm with dephasing times still high $>0.1$ ms~\cite{Ohno2012}. For example,
at $z_{\mathrm{NV}}=20$ nm, the dipolar
interaction frequency becomes $d/2\pi \sim 3.25$ kHz. Closeness to the surface
of the YIG strip is critical to be able to have robust SSE in the presence
of private (local) nuclear spin noises in the NV center hosts. Hence, in our
simulations we consider $5-20$ nm heights. We consider a chain
of $N=1000$ sites which corresponds to a chain of length $L=(N-1)a\approx Na\sim 1.24\,\mu$m. This allows us to consider SSE in the range of $\sim 1\mu$m.
A summary of the parameters is presented in a table at the Appendix \ref{sec:appendix-parameters}.

\subsection{Resonance Condition}\label{sec:resonance}

Let's start by writing the resonance
condition between the magnon mode (Eq.~(\ref{eq:MagnonDispersion})) at $k=0$ and 
an NV center qubit (Eq.~(\ref{eq:dressedNVfreq})) at
location $x_i$ on the chain subject to a bias magnetic field $B_0$
\begin{eqnarray}
\omega(k=0)=\omega_0 = (\omega_i^2+8s\alpha_i^2)^{1/2}.
\end{eqnarray}
Here, $\alpha_i$ and $\beta_i$ (in $\omega_i$ of Eq.~(\ref{eq:omega_i})) are fixed by $x_i$. Both sides
of the resonance condition depend on $B_0$ through 
$\omega_\mathrm{NV}=D-\gamma_\mathrm{NV}B_0$ and $\omega_0=\gamma_0B_0$ (Note that
$\gamma_\mathrm{NV}\approx \gamma_0$). We remark that this resonance should not be confused with the usual ferromagnetic resonance condition,
where time-dependent external fields are involved. Here we only have
static external fields yielding Larmor frequencies. In our case, $k=0$ spin wave mode frequency is matched to the qubit transition frequency.
We numerically solve the implicit equation and find $B_0\sim 51$ mT for $x_1=\pm L/4$.
We numerically verified that resonance condition is weakly dependent on the spatial location of the NV qubits on the YIG nanostrip, unless they are almost exactly at the ends. 
Though, we will limit our discussions to the pairwise entanglement of about half micrometer separated qubits in this
paper, due to the approximately spatially uniform behavior of interaction coefficients and
the resonance condition,
our scheme could be scaled to more NV center qubits straightforwardly. 

\subsection{Spatial profile of coherence function of the bath modes}

From Eq.~(\ref{eq:coh_k}), we can write the coherence function of the magnon
bath modes explicitly
\begin{eqnarray}\label{eq:coh_profile}
\epsilon_k=-i\frac{B_1}{B_0}\sqrt{\frac{s}{2N}}\sum_{j=-N/2}^{N/2} {\cal B}_{1j}\text{e}^{-ik x_j},
\end{eqnarray}
where we express the inhomogeneous external field $\boldsymbol{B_1}$ as $\boldsymbol{B_1}(x_j)=\boldsymbol{B_1}{\cal B}_{1j}$ with ${\cal B}_{1j}\equiv {\cal B}_{1}(x_j)$ 
is a unit amplitude spatial profile function. Coherence function directly contributes to the bath correlation functions through Eqs.~(\ref{eq:DisplacedThermalBath-m})-(\ref{eq:DisplacedThermalBath-mdagm}). Thus, if the spatial space profile of the coherence is too broad, or if the ${\cal B}_{1}(x_j)$ is close to uniform, only the lowest wavelength bath modes would dominate the open system dynamics, making it non-Markovian. While we can externally
control the amount of coherence via the ratio of magnetic field amplitudes 
$B_1/B_0$, the inhomogeneity of $\boldsymbol{B_1}$ can be used to continuously tune non-Markovian character of the magnon bath. 

\begin{figure}[t!]%
    \centering
    \subfloat[\centering \label{fig:fig4-CoherenceProfile}]{{\includegraphics[width=0.47\linewidth]{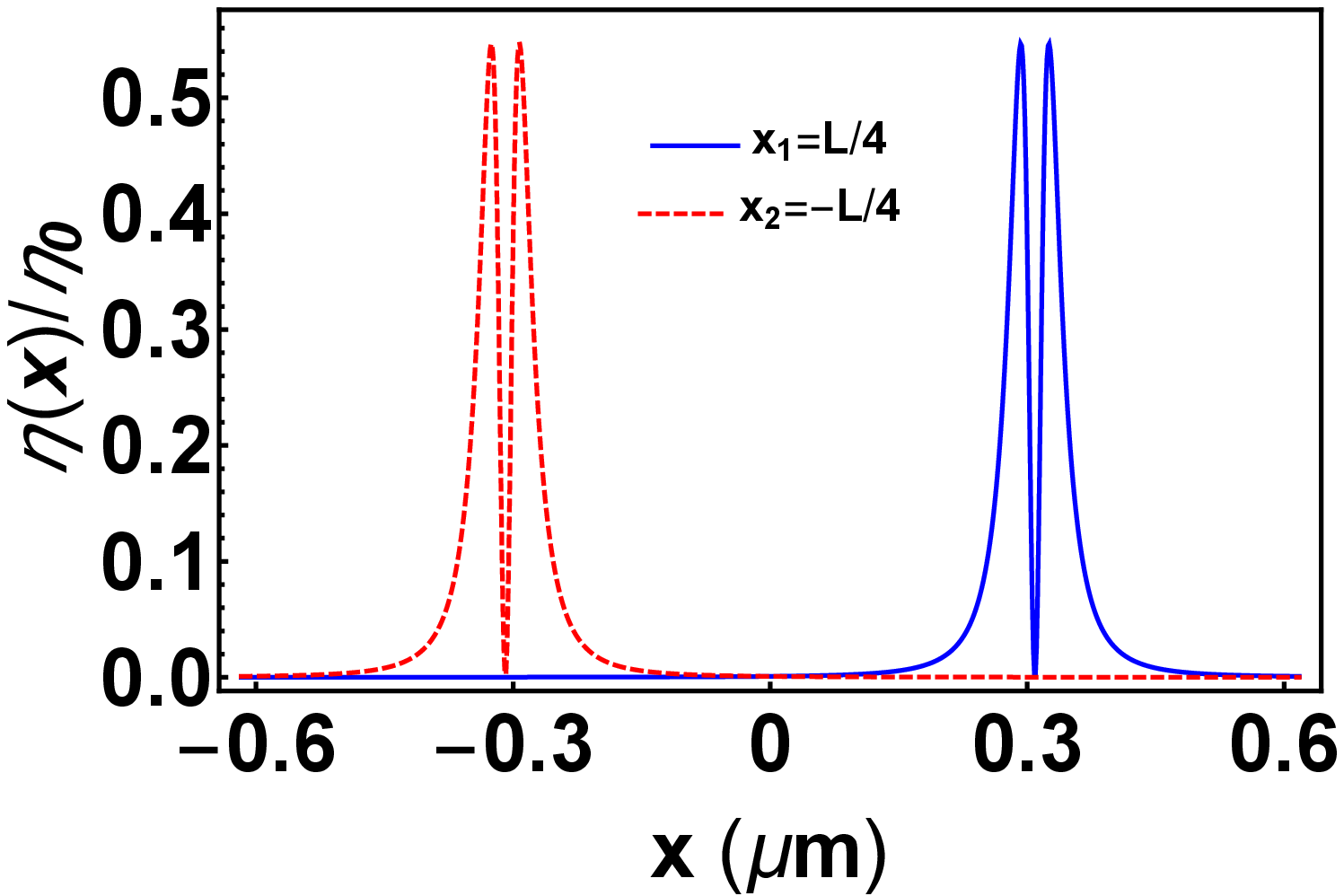}}}\subfloat[\centering \label{fig:etak}]{{\includegraphics[width=0.51\linewidth]{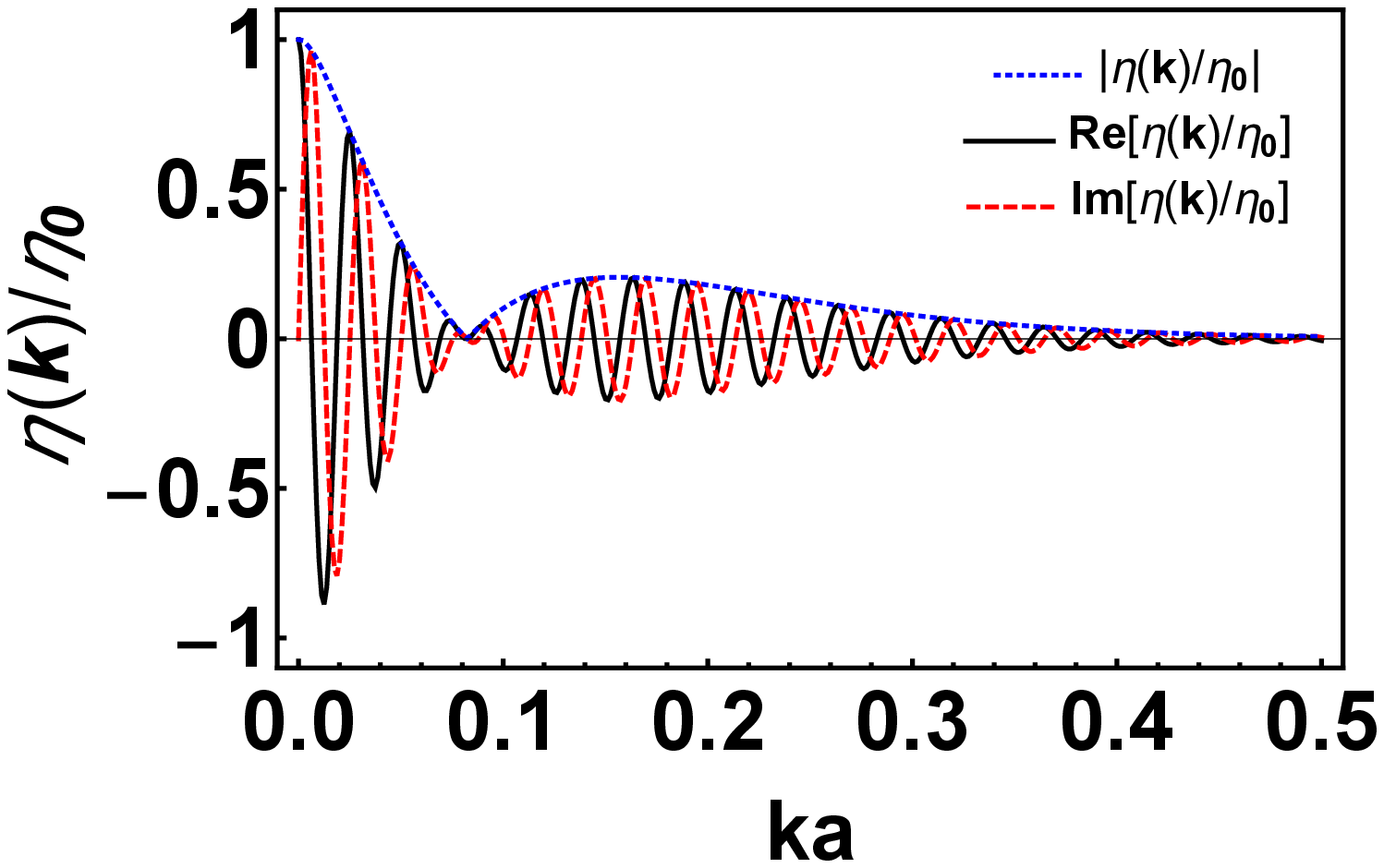} }}%
    \caption{ (Color online) a) Spatial profiles of the interaction coefficients $\eta(x)$ of a pair of NV center qubits placed at $x=L/4$ (blue solid curve) and $x=-L/4$ (red dashed curve) on a YIG strip of length $L\sim 1.2\,\mu$m. The interaction coefficient is normalized with the $\eta_0\sim 5.55165\times 10^4$ Hz, the interaction strength of the $k=0$ magnon mode with the NV center qubits. Both spatial profiles yield the same interaction strength $\eta(k)$ in the reciprocal space. b) Real (solid black curve) and imaginary (red dashed curve) parts, and the norm (dotted blue curve) of the coefficient $\eta(k)$ of the rotating terms in the magnon-NV center qubit interaction as a function of the wavenumber of the magnon mode $k$. NV center qubit is placed at $x=L/4$ from the end of the YIG nanostrip of length $L$. The interaction coefficient is normalized with the $\eta_0\sim 5.55165\times 10^4$ Hz, the interaction strength of the $k=0$ magnon mode with the NV center qubits. $k$ is multiplied with the lattice constant $a$ so that the horizontal axis is dimensionless. $\eta(k)$ is an even function of $k$ and only $k>0$ behavior is shown. 	}
\end{figure}

Our objective is to find simple and intuitive Markovian relaxation towards robust steady-state entanglement, and hence
it is necessary for us to consider sufficiently focused, spatially narrow, $\boldsymbol{B_1}$.
For that aim, and to make the number of parameters in our theory unchanged, we simply
assume ${\cal B}_{1}(x_j)$ has the same spatial behavior as 
$\eta_{ij}\equiv\eta_{ij}$. We show that $\eta_k$ is the significant interaction for the open system relaxation of NV qubits and other interaction
coefficients have similar time scales as with the bath correlation function
determined through $|\eta_k|^2$. Coherence function would bring additional
correlation functions that depend on $\eta_k\epsilon_k$, which we want to be
broad. The spatial profile we take here is only an example and is not a prerequisite in any experimental implementation. One can use different spatial profiles than the one we consider here, provided that $\eta_k\epsilon_k$ is broad enough to give decaying bath
correlations within Markovian time scales. Beyond Markovian regime, our theory is not applicable, but one could explore non-Markovian effects on SSE and SSC by using spatially broader magnetic fields.  Our choice allows for a simple test of Markov
approximation without additional parameters. Plots of the $\eta(x)/\eta_0$ are given in Fig.~\ref{fig:fig4-CoherenceProfile} for two NV center locations $\pm L/4$. Either of them can be taken for ${\cal B}_{1}(x_j)$ as both yield the same coherence
distribution over the modes in $k$-space 
(same as $\eta(k)$, cf.~Fig.~\ref{fig:etak}) so that
\begin{eqnarray}\label{eq:eps}
\epsilon_k=-i\frac{B_1}{B_0}\sqrt{\frac{s}{2N}}\frac{\eta_{k}}{\eta_{0}}
\equiv -i\epsilon\frac{\eta_{k}}{\eta_{0}}.
\end{eqnarray}
We introduced $\epsilon:=(B_1/B_0)(s/2N)^{(1/2)}$ as our coherence parameter 
controlled by the applied magnetic field magnitudes. 
Together with specification of the coherence function and the resonance condition,
we can now develop a Markovian master equation for our system.

\subsection{Master equation for NV centers in a common bath of displaced thermal
	magnons}
Derivation of master equation requires a series of assumptions, which is not
trivial in the case of coherently displaced thermal reservoir and the literature
or the textbooks focus on the case of squeezed bath. Hence, we will start from
the very beginning to see where the assumptions are needed and how they can be
justified. Explicit justification of the so-called Born-Markov approximations is presented in Appendix~\ref{sec:appendix-BornMarkov}.

\captionsetup{labelfont=bf,font=small}

Typically system-bath interactions are much slower
than the free Hamiltonian evolutions and hence it is preferable to use the
interaction picture to follow the interaction dynamics. Writing 
$\hat{H}_0=\hat{H}_{\text{mag}}+\hat{H}_{\text{NV}}$ with 
Eqs.~(\ref{eq:HnvNewBasis}) and~(\ref{eq:HmagDis}) in the unitary $\hat{U}(t,0)=\exp(i\hat{H}_0t/\hbar)$, the interaction picture transformations 
for the overall state $\rho_{SB}$ and the Hamiltonian $\hat{H}_{SB}$ are
given by
\begin{eqnarray}
\rho_{SB}^I(t)&=&\hat{U}(t,0)\rho_{SB}\hat{U}^\dag(t,0),\nonumber\\
\hat{H}_{\text{SB}}^I(t)&=&\hat{U}(t,0)\hat{H}_{SB}\hat{U}^\dag(t,0),
\end{eqnarray}
where $\hat{H}_{\text{SB}}(t)=\hat{H}_\text{deph}(t)+\hat{H}_\text{crt}(t)+
\hat{H}_\text{rt}(t)$ is the overall interaction Hamiltonian and $\rho_{SB}$
is the state of the total system. For brevity we drop the superscript 
$I$ and use only interaction picture operators in what follows.

Infinitesimal time evolution of the overall system under $\hat{H}_{\text{SB}}$, which is given by 
\begin{eqnarray}
\hat U(t+dt,t):=\mathds{1}-\frac{i\hat{H}_{\text{SB}}(t)dt}{\hbar},
\end{eqnarray}
can be applied over a finite time interval $[t,t+\Delta t]$ using the Dyson series in time ordered ($t\ge t_1\ge t_2\ge ...\ge t_n$) manner~\cite{Sakurai2017},
\begin{eqnarray}
\hat{U}&(t+\Delta t,t)=\mathds{1}-\sum_{n=1}^{\infty}\left(\frac{-i}{\hbar}\right)^n
\int_t^{t+\Delta t}dt_1\int_t^{t_1}dt_2...\nonumber\\
&...\int_t^{t_{n-1}}dt_n
\hat{H}_{\text{SB}}(t_1)\hat{H}_{\text{SB}}(t_2)...\hat{H}_{\text{SB}}(t_n).
\end{eqnarray}
If the system-bath coupling is weaker relative to the free evolution, we can terminate the Dyson series after the second order. Even when the leading first order
term is non-vanishing, the second order term is kept as it is responsible
to describe irreversible system dynamics in an environment. Substituting the terminated $\hat{U}(t+\Delta t,t)$ into the 
$\rho_{\mathrm{SB}}(t+\Delta t)=\hat{U}(t+\Delta t,t)\rho_{\mathrm{SB}}(t)\hat{U}^\dag(t+\Delta t,t)$, and using
\begin{eqnarray}
\int_t^{t+\Delta t}&dt_1&\int_t^{t+\Delta t}dt_2\hat A(t_1)A(t_2)=\nonumber\\
&2&\int_t^{t+\Delta t}dt_1\int_t^{t_1}dt_2A(t_1)A(t_2),
\end{eqnarray}
for any operator $\hat A(t)$, and
\begin{eqnarray}
\int_t^{t+\Delta t}&dt_1&\int_t^{t_1}dt_2A(t_1)A(t_2)=\nonumber\\
&&\int_t^{t+\Delta t}dt_1\int_0^{\Delta t}dsA(t_1)A(t_1-s),
\end{eqnarray}
we find
\begin{align}\label{eq:BlochRedfield1}
\begin{split}
&\rho_{\text{SB}}(t+\Delta t)-\rho_{\text{SB}}(t)=\frac{-i}{\hbar}\int_t^{t+\Delta t}dt_1
[\hat{H}_{\text{SB}}(t_1),\rho_{\text{SB}}(t)]\nonumber\\
&-\frac{1}{\hbar^2}\int_t^{t+\Delta t}dt_1\int_0^{\Delta t}ds
[\hat{H}_{\text{SB}}(t_1),[\hat{H}_{\text{SB}}(t_1-s),\rho_{\text{SB}}(t)]
\end{split}\\
\end{align}

Let's suppose that the bath has 
many degrees of freedom (modes), yielding a broad, continuous bath spectrum.
Accordingly, the bath dynamics can be treated independently, and its
equilibrium state can be taken as the initial bath state $\rho_B$, which 
cannot change significantly under the weak
system-bath coupling. The system-bath state factorization 
$\rho_{SB}(t)\approx \rho(t)\otimes \rho_B(t)$ and frozen initial bath state
$\rho_B=\rho_B(t)$
assumptions are known as the Born approximations~\cite{Breuer2007}.

After tracing out the bath degrees of freedom we get the irreversible dynamics of the
system, whose characteristic time scale is denoted by $\tau_s$. 
If we take $\Delta t\ll \tau_s$, $\Delta t$ becomes a coarse-grained, effectively
infinitesimal, time step for the system dynamics. The integrals over $dt_1$ are simplified to $\Delta t$
and the $t_1$ dependent integrands are evaluated at $t_1=t$. Dividing the equation
by $\Delta t$, the left hand side can be replaced by a coarse-grained differential
$\rho_{\mathrm{SB}}(t+\Delta t)-\rho_{\mathrm{SB}}(t))/\Delta t\equiv{d\rho(t)/dt}$.
If we write the system-bath interaction in a generic form
$\hat{H}_{\mathrm{SB}}(t)= \hat S_k\otimes\hat B_k$, where summation over repeated
index is implied, one can see that the integrands include the so-called two-time 
bath correlation functions 
$G_{kl}(t,t-s)=\langle \hat B_k(t)\hat B_l(t-s) \rangle
=\mathrm{Tr}[\rho_B \hat B_k(t)\hat B_l(t-s)]$. If these bath correlators decay
significantly in a time $\tau_B$ that lies within the coarse-grained time step $\Delta t$, then $\Delta t$ in the
upper limit of the remaining integral over $s$ can be replaced by $\infty$.
The hierarchy of the time scales $\tau_B<\Delta t<\tau_s$ and the associated manipulations of the integral expressions are known as 
Markov approximations~\cite{Breuer2007}. It is necessary for us to determine
the time scales self-consistently by specifying our physical system and
the corresponding parameters, which is the subject of subsequent sections.
Here, we continue with stating the final expression after the Born-Markov
approximations, also known as the Born-Markov master equation
\begin{eqnarray}\label{eq:masterEqGeneral}
&\dot\rho(t)=
i\Tr_B[\rho_{SB}(t),\hat{H}_{\text{SB}}(t)]
+{\cal L}\rho(t),
\end{eqnarray}
where the Liouvillian superoperator ${\cal L}$ is defined to be
\begin{eqnarray}\label{eq:mapL}
{\cal L}\rho=\Tr_B\int_0^{\infty}ds
[\hat{H}_{\text{SB}}(t),[\rho\otimes\rho_B,\hat{H}_{\text{SB}}(t-s)]].
\end{eqnarray}
Here and in what follows, we drop the factors of $1/\hbar$ and  $1/\hbar^2$, assuming that $\hat{H}_{\mathrm{SB}}$ and all the other Hamiltonians
are scaled with $\hbar$.

To continue with the calculation of the master equation, a compact
expression of $\hat{H}_{\mathrm{SB}}(t)$ is convenient. For that aim,
we introduce the magnon bath operators,
\begin{eqnarray}\label{eq:bathOps}
\hat{B}_i^\alpha(t):=\sum_k (f_k^{i\alpha}\hat{m}_k(t)
+g_k^{i\alpha}\hat{m}_k^\dag(t)),
\end{eqnarray}
where $\hat{m}_k(t)=\hat{m}_k\exp(-i\omega_k t)$ and 
\begin{eqnarray}\label{eq:bathOpsCoeffs}
f_k^{iz}&=&\xi_k^{(i)},\quad g_k^{iz}=\xi_k^{(i)\ast}\nonumber\\
f_k^{i-}&=&\zeta_k^{(i)},\quad g_k^{i-}=\eta_k^{(i)\ast}\\
f_k^{i+}&=&\eta_k^{(i)},\quad g_k^{i+}=\zeta_k^{(i)\ast}.\nonumber
\end{eqnarray}
In addition, the interaction picture operators of the qubits will be denoted
by $\hat\sigma_i^\alpha(t)$ such that
\begin{eqnarray}\label{eq:qubitOps}
\hat\sigma_i^\alpha(t)=\hat\sigma_i^\alpha \exp(i\Omega_i^\alpha t),
\end{eqnarray}
where $\alpha\in\{z,\pm\}$, $\Omega_i^z=0$, $\Omega_i^\pm = \pm\Omega_i$, and $\hat\sigma_i^-\equiv\hat\sigma_i$. In terms of these short-hand notations,
the interaction Hamiltonian is expressed as
\begin{eqnarray}\label{eq:HsbCompact}
\hat{H}_{\text{SB}}(t)=
\sum_{i\alpha}\hat\sigma_i^\alpha(t)\hat{B}_i^\alpha(t).
\end{eqnarray}

After the substitution of the Hamiltonian (\ref{eq:HsbCompact}), 
the first term  of the master equation (\ref{eq:masterEqGeneral})
can be expressed in a Liouville-von Neumann form
$i[\rho(t),H_\mathrm{drive}]$ in terms of the effective driving
Hamiltonian,
\begin{eqnarray}\label{eq:HeffDrive}
\hat{H}_{\text{drive}}(t)=
\sum_{i\alpha}\hat\sigma_i^\alpha(t)\langle \hat{B}_i^\alpha(t)\rangle.
\end{eqnarray}
This term can only contribute when coherence is injected to the magnons with $\boldsymbol{B_1}$.

If the qubits are placed symmetrically about the center of the linear chain, 
the interaction coefficients are the same and we can drop the index $i$ from the
bath operators. Substituting the coherence parameter from Eq.~(\ref{eq:eps}) into
the $\langle \hat{B}_i^\alpha(t)\rangle$, the effective drive term (\ref{eq:HeffDrive}) in the Schr\"odinger picture becomes
\begin{eqnarray}\label{eq:HeffDrive2}
\hat{H}_{\text{drive}}(t)=&-&\sum_{ik}\hat\sigma_i^-[\zeta_{k}\epsilon_{k}
\text{e}^{-i(\omega_0+\omega_{k})t}
+\eta_{k}^\ast\epsilon_{k}^\ast \text{e}^{-i(\omega_0-\omega_{k})t}]
\nonumber\\
&-&\sum_{ik}\hat\sigma_i^+[\eta_{k}\epsilon_{k}
\text{e}^{i(\omega_0-\omega_{k})t}
+\zeta_{k}^\ast\epsilon_{k}^\ast \text{e}^{i(\omega_0+\omega_{k})t}]
\nonumber\\
&-&\sum_{ik}\hat\sigma_i^z[\xi_{k}\epsilon_{k}
\text{e}^{-i\omega_{k}t}
+\xi_{k}^\ast\epsilon_{k}^\ast \text{e}^{i\omega_{k}t}],
\end{eqnarray}
where we have used the resonance condition $\Omega_i=\omega_0$ in time dependence
of the interaction picture qubit operators. We can separate the resonant terms with
$k=0$ from those off-resonant terms with $k\neq 0$ in this Hamiltonian. Dropping these off-resonant terms is equivalent to the employing the rotating wave approximation (RWA)~\cite{scully_zubairy_1997} to every off-resonant term and to keep only the static terms. For a finite length YIG strip this approximation can be justified.
We consider a chain of $N=10^3$ sites, corresponding to $L\sim 1.2\,\mu$m. This gives
a separation between mode frequencies $\omega_k$ in the order of $\sim 0.1\omega_0$,
which is much larger than the interaction coefficients $\eta_k\sim 10^{-5}\omega_0$.
The effective drive Hamiltonian under the RWA in the Schr\"{o}dinger picture then simplifies to
\begin{eqnarray}\label{eq:HeffDrive3}
\hat{H}_{\text{drive}}=-\eta_0\epsilon
\sum_{i}\hat\sigma_i^y,
\end{eqnarray}
where $\hat\sigma_i^y=-i(\hat\sigma_i^+-\hat\sigma_i^-)$.

We expand the commutator in the second term of Eq.~(\ref{eq:masterEqGeneral})
and substitute the Hamiltonian~(\ref{eq:HsbCompact}) which gives the Bloch-Redfield
master equation~\cite{Breuer2007} in the form,
\begin{eqnarray}\label{eq:BlochRedfieldMasterEq}
{\cal L}\rho=\sum_{ij\alpha\beta}
{\text{e}^{i(\Omega_i^\alpha+\Omega_j^\beta)t}}
G_{ij}^{\alpha\beta}(\Omega_j^\beta,t)
[\hat\sigma_j^\beta\rho,\hat\sigma_i^\alpha]+\text{H.c.}.
\end{eqnarray}
We introduced the one-sided Fourier transform of 
two-time bath correlation functions, 
\begin{eqnarray}
G_{ij}^{\alpha\beta}(t-s,t)=
\Tr_B(\hat B_i^\alpha(t)\hat B_j^\beta (t-s)),
\end{eqnarray}
as follows,
\begin{eqnarray}\label{eq:1sidedFourierG}
G_{ij}^{\alpha\beta}(\omega,t)=\int_0^\infty\,ds
{\text{e}^{-i\omega s}}G_{ij}^{\alpha\beta}(t-s,t).
\end{eqnarray}
In contrast to usual derivations of the master equation, the condition of
stationary bath state, $[\rho_B,H_\mathrm{mag}]=0$ is not sufficient to
have temporally homogeneous correlations with 
$G_{ij}^{\alpha\beta}(t-s,t)=G_{ij}^{\alpha\beta}(0,s)$ for our displaced
thermal bath. The integral over $s$ in Eq.~(\ref{eq:1sidedFourierG}) 
can be taken using
\begin{eqnarray}
\int_0^\infty\,ds \text{e}^{\pm i\omega s}=
\pi\delta(\omega)\pm i{\cal P}\left(\frac{1}{\omega}\right),
\end{eqnarray}
where ${\cal P}$ denotes the Cauchy principal value. The second term gives
rise to a small Lamb shift Hamiltonian, which can be neglected relative to
the drive and the free Hamiltonian of the qubits. After the integration, 
$G_{ij}^{\alpha\beta}(\Omega_j^\beta,t)$ becomes
\begin{eqnarray}\label{eq:GsumOverKQ}
G_{ij}^{\alpha\beta}(\Omega_j^\beta,t)
&=&\pi\sum_{kq}\left(f_k^{i\alpha}f^{j\beta}_q
\epsilon_k\epsilon_q\delta(\Omega_j^\beta-\omega_q)
\text{e}^{-i(\omega_k+\omega_q)t}
\right.\nonumber\\
&+&f_k^{i\alpha}g^{j\beta}_q
\epsilon_k\epsilon_q^\ast\delta(\Omega_j^\beta+\omega_q)
\text{e}^{-i(\omega_k-\omega_q)t}\nonumber\\
&+&g_k^{i\alpha}f^{j\beta}_q
\epsilon_k^\ast\epsilon_q\delta(\Omega_j^\beta-\omega_q)
\text{e}^{-i(\omega_k-\omega_q)t}\nonumber\\
&+&\left. g_k^{i\alpha}g^{j\beta}_q
\epsilon_k^\ast\epsilon_q^\ast\delta(\Omega_j^\beta+\omega_q)
\text{e}^{-i(\omega_k+\omega_q)t}\right).\nonumber\\
&+&\pi\sum_{k}\left(f_k^{i\alpha}g^{j\beta}_k
(\bar{n}_k+1)\delta(\Omega_j^\beta+\omega_k)
\right.\nonumber\\
&+&\left. g_k^{i\alpha}f^{j\beta}_k
\bar{n}_k\delta(\Omega_j^\beta-\omega_k)
\right)
\end{eqnarray}

The resonance condition $\Omega_i^\beta=\omega_0$ fixes the $\beta=\pm$ and $q=0$ in the first four terms of 
Eq.~(\ref{eq:GsumOverKQ}), after converting the summation over $q$ to an integral over $\omega_q$. Similarly, we replace the summation over $k$ with an integral
over $\omega_k$ in the last two terms. Effectively, we can use the replacements in each term
\begin{eqnarray}\label{eq:deltaFunctions}
\delta(\Omega_j^\beta\mp\omega_p)\rightarrow\frac{D_0}{2\pi}
\delta_{\beta \pm}\delta_{p0}
\end{eqnarray}
with $p\in\{k,q\}$. This lefts us two forms of summations over $k$ in the first four terms,
\begin{eqnarray}
S_1&:=&\sum_k f_k^{i\alpha}\epsilon_k\text{e}^{-i\omega_k t},\\
S_2&:=&\sum_k g_k^{i\alpha}\epsilon_k^\ast\text{e}^{i\omega_k t}.
\end{eqnarray}
They can be controlled by the spatial profile of the magnetic field $\bm{B_1}$.
$S_1$ and $S_2$ explicitly become, after substitution of the $\epsilon_k$ of Eq.~(\ref{eq:eps}), 
\begin{eqnarray}
S_1&:=&-i\frac{\epsilon}{\eta_{0}}
\sum_k f_k^{i\alpha}\eta_{k}\text{e}^{-i\omega_k t},\label{eq:S1}\\
S_2&:=&-i\frac{\epsilon}{\eta_{0}}
\sum_k g_k^{i\alpha}\eta_{k}^\ast\text{e}^{i\omega_k t}.\label{eq:S2}
\end{eqnarray}

The off-resonant terms in the master equation oscillating at such a high frequency can still be regarded 
as fast relative to the static (resonance) terms, as we argued in the RWA for the 
drive term, and they can be dropped in the dissipator terms, too,
according to the full secular approximation~\cite{Cattaneo2019}. In a more rigorous partial secular approximation, some time dependent terms are kept in such a way that the dynamical hierarchy of 
dissipation terms are respected~\cite{Cattaneo2019}. Partial secular 
approximation keeps the operator structure of the master equation same as the
full secular approximation. Additional time dependent oscillatory shifts to
the dissipation rates emerge, which can bring qualitative (oscillatory) changes 
in the dynamics. As our focus is on steady state behavior, we employ the full secular
approximation here.

Substitution of Eqs.~(\ref{eq:deltaFunctions}) and~(\ref{eq:S1})-(\ref{eq:S1}) into 
Eq.~(\ref{eq:GsumOverKQ}) gives a long expression for $G_{ij}^{\alpha\beta}(\Omega_j^\beta,t)$, which is simplified after multiplication with $\exp{[i(\Omega_{i}^\alpha+\Omega_j^\beta)t]}$ and application of the full
secular approximation to 
\begin{eqnarray}
\text{e}^{i(\Omega_i^\alpha+\Omega_j^\beta)t}
&G&_{ij}^{\alpha\beta}(\Omega_j^\beta,t)\approx
-\frac{\kappa}{2}\epsilon^{2}\delta_{\alpha+}\delta_{\beta+}
\nonumber\\
&+&\frac{\kappa}{2}(\bar{n}_0+1+\epsilon^{2})\delta_{\alpha+}\delta_{\beta-}\nonumber\\
&+&\frac{\kappa}{2}(\bar{n}_0+\epsilon^{2})\delta_{\alpha-}\delta_{\beta+}\nonumber\\
&-&\frac{\kappa}{2}\epsilon^{2}\delta_{\alpha-}\delta_{\beta-}.
\end{eqnarray}
Here, we introduced $\kappa=D_0\eta_0^2$.

The Bloch-Redfield master equation~(\ref{eq:BlochRedfieldMasterEq}) in
the Schr\"odinger picture becomes
\begin{eqnarray}\label{eq:masterEqFinal}
\frac{d\rho(t)}{dt}&=&
\frac{i}{\hbar}[\rho,\hat{H}_{\text{NV}}(t)+\hat{H}_{\text{drive}}]
\nonumber\\
&-&\frac{\kappa\epsilon^{2}}{2}\sum_{ij}[
D(\hat\sigma_i^+,\hat\sigma_j^+)
+D(\hat\sigma_i,\hat\sigma_j)]\nonumber\\
&+&\frac{\kappa}{2}(\bar{n}_0(T)+1+\epsilon^{2})
\sum_{ij}D(\hat\sigma_i,\hat\sigma_j^\dag)
\nonumber\\
&+&\frac{\kappa}{2}(\bar{n}_0(T)+\epsilon^{2})
\sum_{ij}D(\hat\sigma_i^\dag,\hat\sigma_j)
\end{eqnarray}
The dissipator superoperators are written in the form
\begin{eqnarray}
D(A,B):=(2A \rho B - \{BA,\rho\}).
\end{eqnarray}

Liouvillian superoperator is traceless and
hence the master equation is governed by a trace preserving map. 
The first term in the Liouvillian is not in the  GKLS 
(Gorini, Kossakowski, Lindblad, Sudarshan) form, hence it is not immediately 
obvious that the evolution described by such a map is completely positive.
The master equation we obtained however is identical with that of open
system dynamics in a squeezed thermal reservoir. Complete positivity and trace preserving (CPTP) conditions are satisfied
by squeezed thermal bath master equation as it can be brought into manifestly
GKLS form using atomic Bogoluibov transformations~\cite{Banerjee_2007}.

The absorption and emission dissipators in the master equation include non-local terms that couple different qubits. While a common thermal bath can be sufficient for generating SSE of initially uncorrelated qubits,
such an entanglement can be fragile in the presence other decoherence channels. 
In addition to magnon bath, the NV center qubits are subject to
their private nuclear spin environments ($^13$C nuclear spins) in the diamond hosts, which cause additional 
dephasing and decoherence. They contribute to the master equation with the Liouvillian
\begin{eqnarray}\label{eq:Liouvillian-NV}
{\cal L}_\text{NV}\rho&=&
\frac{\kappa_\text{NV}}{2}(
(\bar{n}_0(T)+1)D(\hat\sigma_i,\hat\sigma_i^\dag)
+\bar{n}_0(T)D(\hat\sigma_i^\dag,\hat\sigma_i)
)\nonumber\\
&+&\frac{\kappa_{\text{NV}}^{\text{deph}}}{2}
(\hat\sigma_i^z\rho\hat\sigma_i^z-\rho),
\end{eqnarray}
where $\kappa_\mathrm{NV}$ and $\kappa_{\mathrm{NV}}^{\mathrm{deph}}$ denote the dissipation and dephasing rates of the NV centers to their local nuclear spin baths, respectively. We assume the same rates for each qubit for simplicity. In terms of the longitudinal relaxation (dissipation or equilibrium) time $T_1$ and transverse relaxation (dephasing) time we can write $\kappa_\mathrm{NV}=1/T_1$ and $\kappa_{\mathrm{NV}}^{\mathrm{deph}}=1/T_2$. Using cryogenic cooling to $\sim 77$ K
and dynamical decoupling techniques, $T_2\approx 0.6$ s can be achieved~\cite{Bar-Gill2013}). At higher
temperatures available with thermoelectric cooling ($> 160$ K), dephasing get faster
with $T_2\approx 40$ ms~\cite{Bar-Gill2013,Barry2020}). With the theoretical relation for two-level systems $T_2=2T_1$ (In practice, depending on
the settings and the methods one could get different relations such as $T_2=0.5T_1$~\cite{Bar-Gill2013}), same order of longitudinal relaxation time can
be expected. Accordingly,
for the ultralow temperature regimes we consider $T_1$ can be several hours~\cite{Astner2018}, while at low, cryogenic temperatures, relaxation times of tens of seconds are possible. Therefore
we could neglect the local 
dephasing and dissipation of the NV centers to their nuclear spin baths described by ${\cal L}_\mathrm{NV}$. On the other hand, dynamical decoupling methods are energetically costly. We aim to see how robust our scheme is without using such additional methods, and therefore we will systematically examine the effects of $T_1$ and $T_2$ on the entanglement dynamics in the range of milliseconds to seconds. Moreover, we point out in the next section that there are surprising beneficial effects of local dissipation to enhance SSE and SSC, too. In addition, there can be another and more severe decoherence channel unique to nanodiamonds due to the surface spins~\cite{Song2014}. For spherical
nanodiamonds, it is found that $T_2\sim 3\,\mu$s for radius of $20$ nm. On the other hand, very recent studies reveal that at ultralow temperatures nanodiamonds of
size $\sim 20$ nm can have $T_1\sim 0.5$ ms~\cite{DeGuillebon2020}.  
We can envision the surface spin effects could be made negligible on a single NV spin by optimizing the geometric shape of the nanodiamond with the location of the NV spin relative to the crystal surfaces. Alternatively, a bulk NV center with a defect close to one of its surfaces could be used at the cost of degradation of scalability of our scheme.

The dissipators with pairwise emission and absorption 
terms (or so-called squeezing-like terms) contribute further to the coupling of
qubits. Besides, their coherent character can enhance the entanglement, making it
more robust. We therefore consider a displaced thermal bath and treat its
coherence characterized by $\epsilon$ as our main control parameter to get steady- state entanglement in the presence of other decoherence channels. Surprisingly the relation between the coherence of the magnon bath and the entanglement is not monotonic, contrary to what one might expect. We cannot simply increase bath coherence to get entanglement. From the structure of the master equation, we see that coherence contribute to local thermal channels and hence can act as if it is thermal noise as well. Therefore, we expect a competitive character in coherence where it can make entanglement worse or it can enhance it, which suggest that there must be a critical coherence for which the entanglement is optimum. Starting with an example physical system, our final objective is to determine such an optimal pairwise steady-state entanglement of qubits for a critical coherence of their public thermal bath, even under additional private decoherence channels of each qubit.

\begin{figure}[t!]
	\centering
	\subfloat[][]{%
		\includegraphics[width=0.49\linewidth]{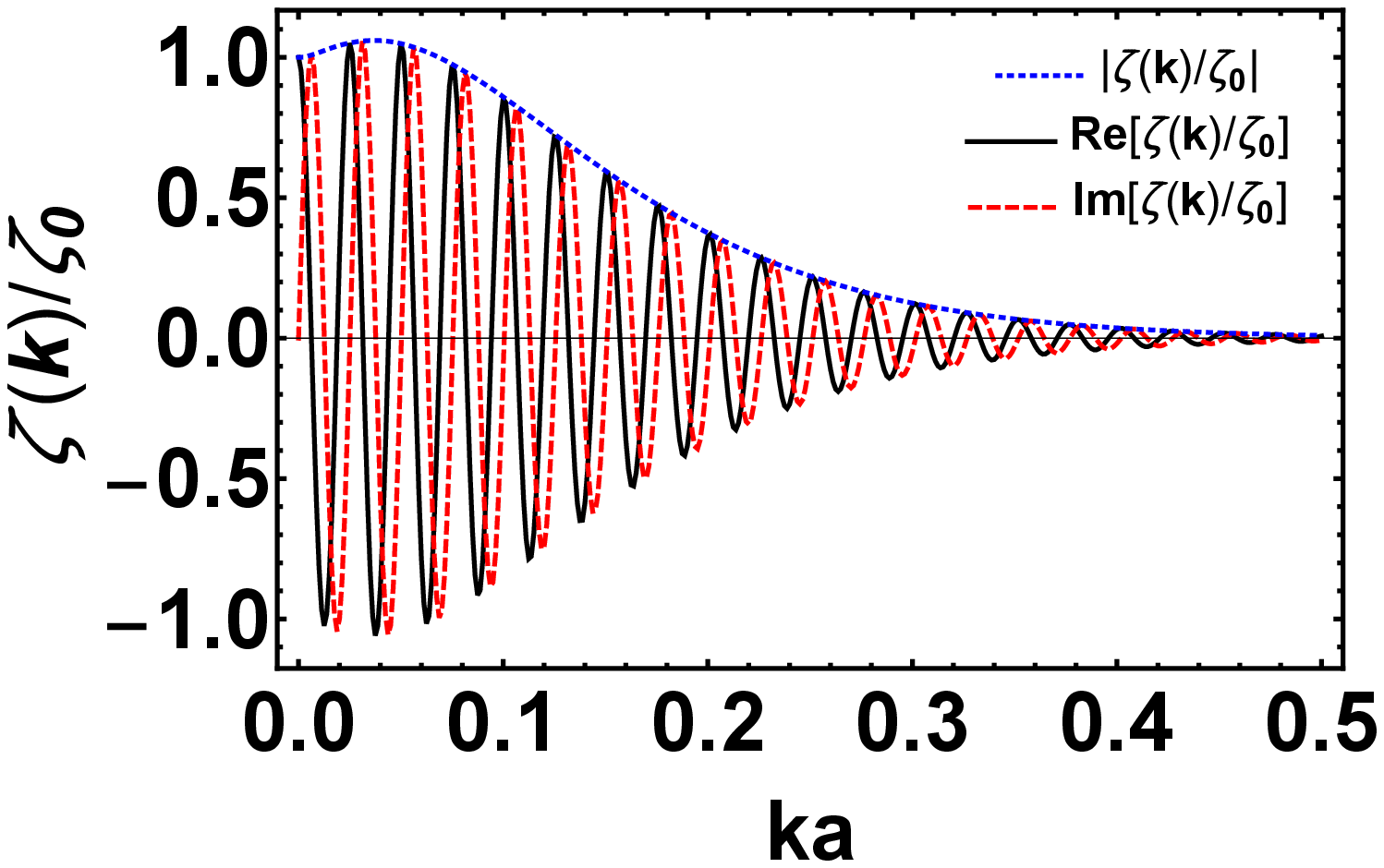}
		\label{fig:zetak}
	}
	\subfloat[][]{%
		\includegraphics[width=0.49\linewidth]{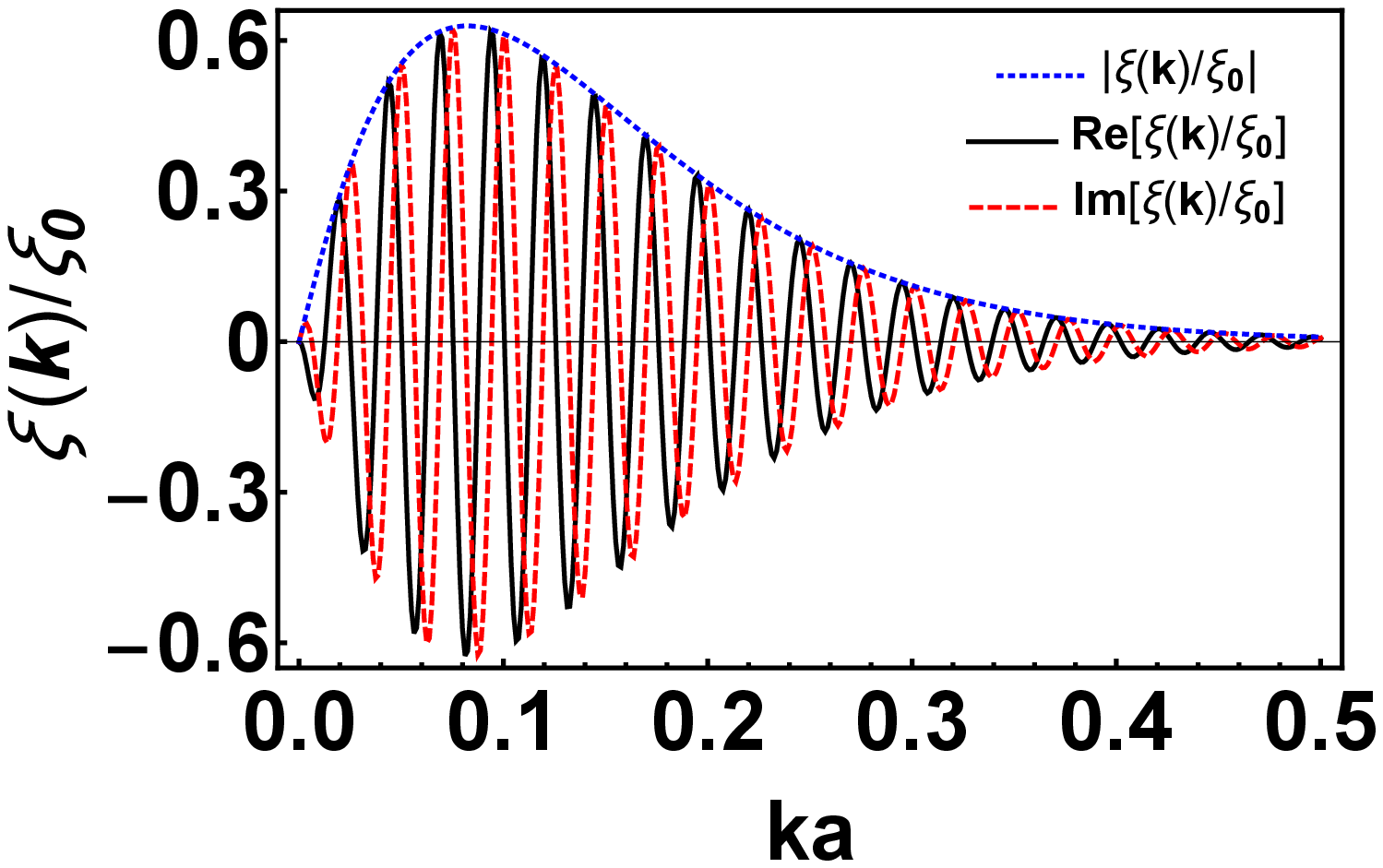}
		\label{fig:xik}
	}
	\caption{(Color online) (a) Real (solid black curve) and imaginary (red dashed curve) parts, and the norm (dotted blue curve) of the coefficient $\zeta(k)$ of the counter rotating terms in the magnon-NV center qubit interaction as a function of the wavenumber of the magnon mode $k$. NV center qubit is placed at $x=L/4$ from the end of the YIG nanostrip of length $L$. The interaction coefficient is normalized with the $\zeta_0\sim -5.557725\times 10^4$ Hz, the interaction strength of the $k=0$ magnon mode with the NV center qubits. $k$ is multiplied with the lattice constant $a$ so that the horizontal axis is dimensionless. $\zeta(k)$ is an even function of $k$ and only $k>0$ behavior is shown. (b) Real (solid black curve) and imaginary (red dashed curve) parts, , and the norm (dotted blue curve)  of the coefficient $\xi(k)$ of the dephasing terms in the magnon-NV center qubit interaction as a function of the wavenumber of the magnon mode $k$. NV center qubit is placed at $x=L/4$ from the end of the YIG nanostrip of length $L$. The interaction coefficient is normalized with the $\xi_0\sim -7.316$ Hz, the interaction strength of the $k=0$ magnon mode with the NV center qubits. $k$ is multiplied with the lattice constant $a$ so that the horizontal axis is dimensionless. $\xi(k)$ is an even function of $k$ and only $k>0$ behavior is shown. 
	}
\end{figure}

\subsection{Steady-state coherence and entanglement}\label{sec:results-SSE}

\subsubsection{NV center qubits in a public magnon bath}\label{sec:SSEnoNVlocalChannels}
When the scaling of the SSE to multiple qubits is not required, one
can consider bulk diamonds or relatively larger and geometrically optimized nanodiamonds to neglect local decoherence channels of the NV centers due to their
nuclear spin environments; in addition dynamical dephasing methods can
be used to eliminate the local dephasing channels.
While this is not energetically efficient case, our objective here is to clarify the physical mechanism of SSE and SSC. Besides to see if any different roles the control parameters can play to get SSE and SSC when there is only public bath and when there are additional private baths. 

\begin{figure}[t!]
	\centering
	\subfloat[][]{%
		\includegraphics[width=0.23\textwidth]{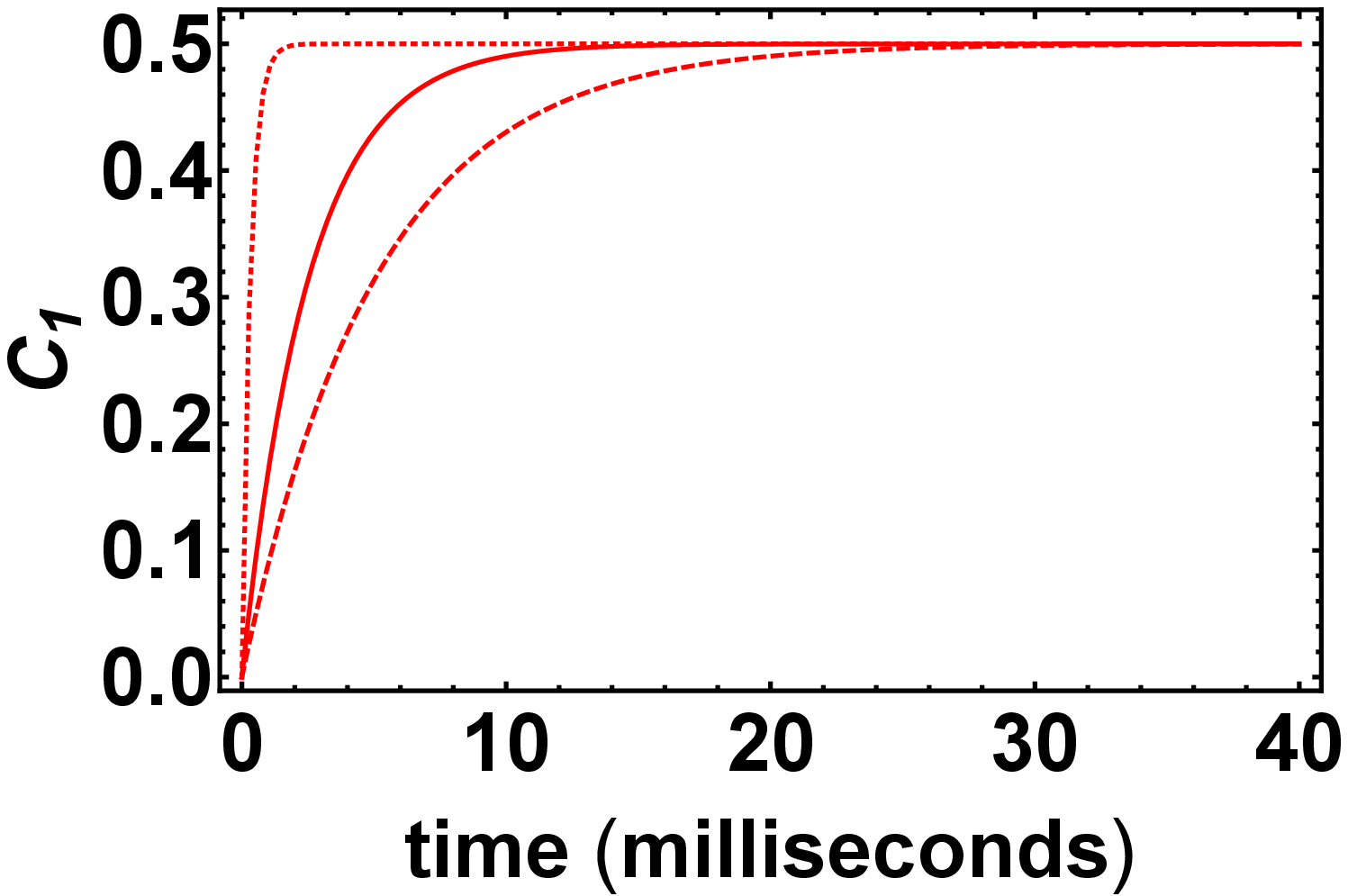}
		\label{fig:CohVsGeo}
	}
	\subfloat[][]{%
		\includegraphics[width=0.23\textwidth]{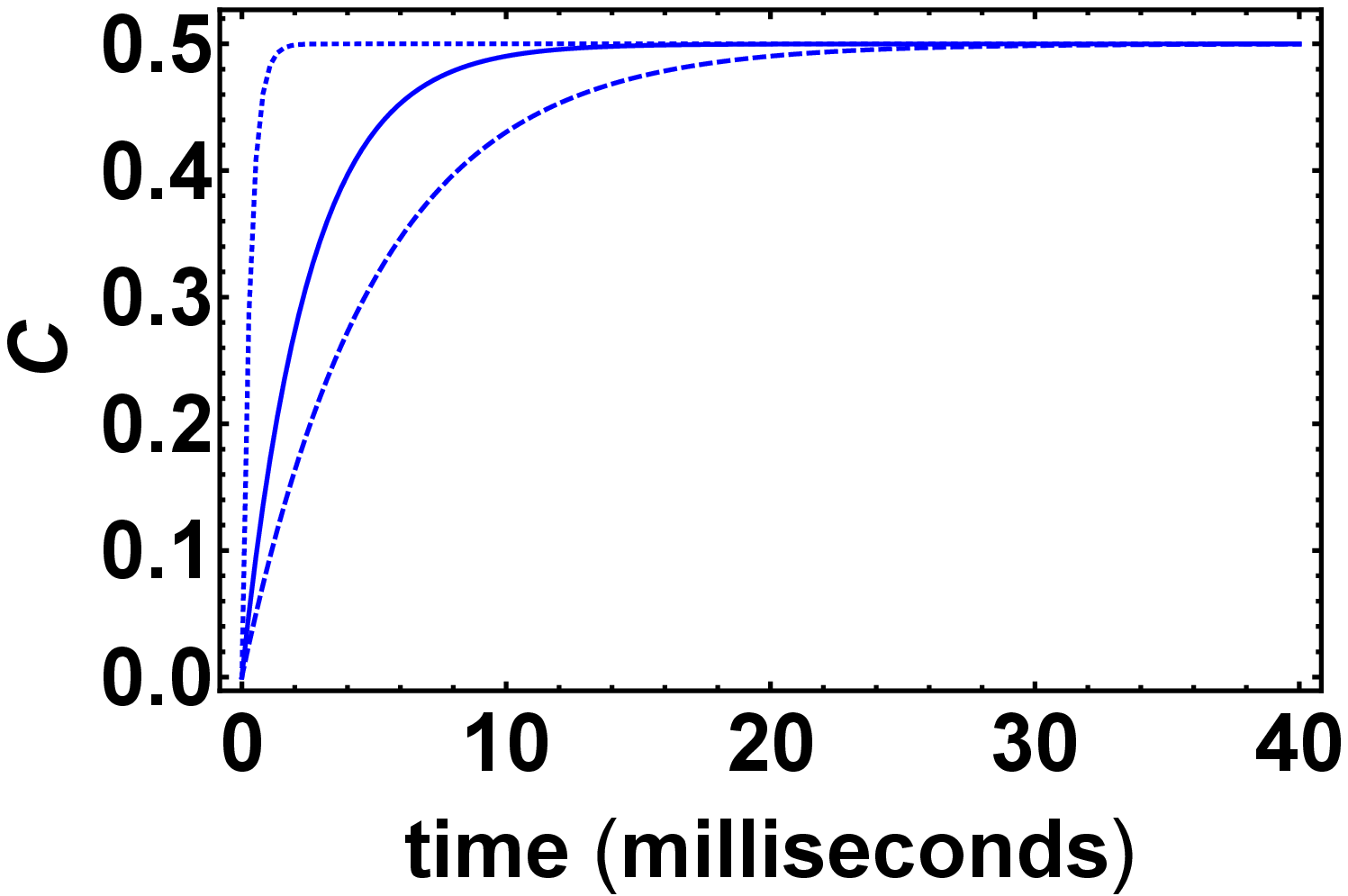}
		\label{fig:ConcVsGeo}
	}
	\caption{\label{fig:CorrVsGeo} (Color online) Dynamics of the $l_1$-norm coherence $C_1$ (red curves in (a)) and Concurrence $C$ (blue curves in (b)) of two NV center qubits in a quasi-one-dimensional thermal magnon bath, for geometric parameters $L_z=10$ nm, $z_{\text{NV}}=20$ nm ( solid red and blue curves), $L_z=20$ nm, $z_{\text{NV}}=20$ nm (dashed red and blue curves), and $L_z=20$ nm, $z_{\text{NV}}=10$ nm (dotted red and blue curves). The other parameters are $\epsilon= 0$, $E=0$, $L_x=1.236\,\mu$m, $T=1$ mK, $T_1,T_2^\ast\rightarrow\infty$ s, $x_{1,2}=\pm L_x/4$ m.}
\end{figure}

Our main geometrical parameters are the thickness of the YIG strip $L_z$ and the height of the NV center qubits from the strip $z_\mathrm{NV}$. The relative locations of NV centers are also of little influence unless they are too close to the ends.
Fig.~\ref{fig:CorrVsGeo} shows that the smaller the $L_z$ or $z_\mathrm{NV}$, the faster SSE is reached, but the amount of SSE and SSC remains the same. In particular, due to the short range nature of the dipole interaction, speed of reaching the steady-state is most sensitive to $z_{\mathrm{NV}}$. We conclude that thinner YIG waveguides and especially NV centers closer to the surface offer faster SSE, which can be beneficial against private nuclear spin noises. Remarkably, the electric field belongs to the geometrical set of parameters in our model as its role is reduced to decreasing the $L_z$ effectively by an electrical length $L_E$ introduced in Eq.~(\ref{eq:dosStripE}).

\begin{figure}[t!]
	\centering
	\subfloat[][]{%
		\includegraphics[width=0.23\textwidth]{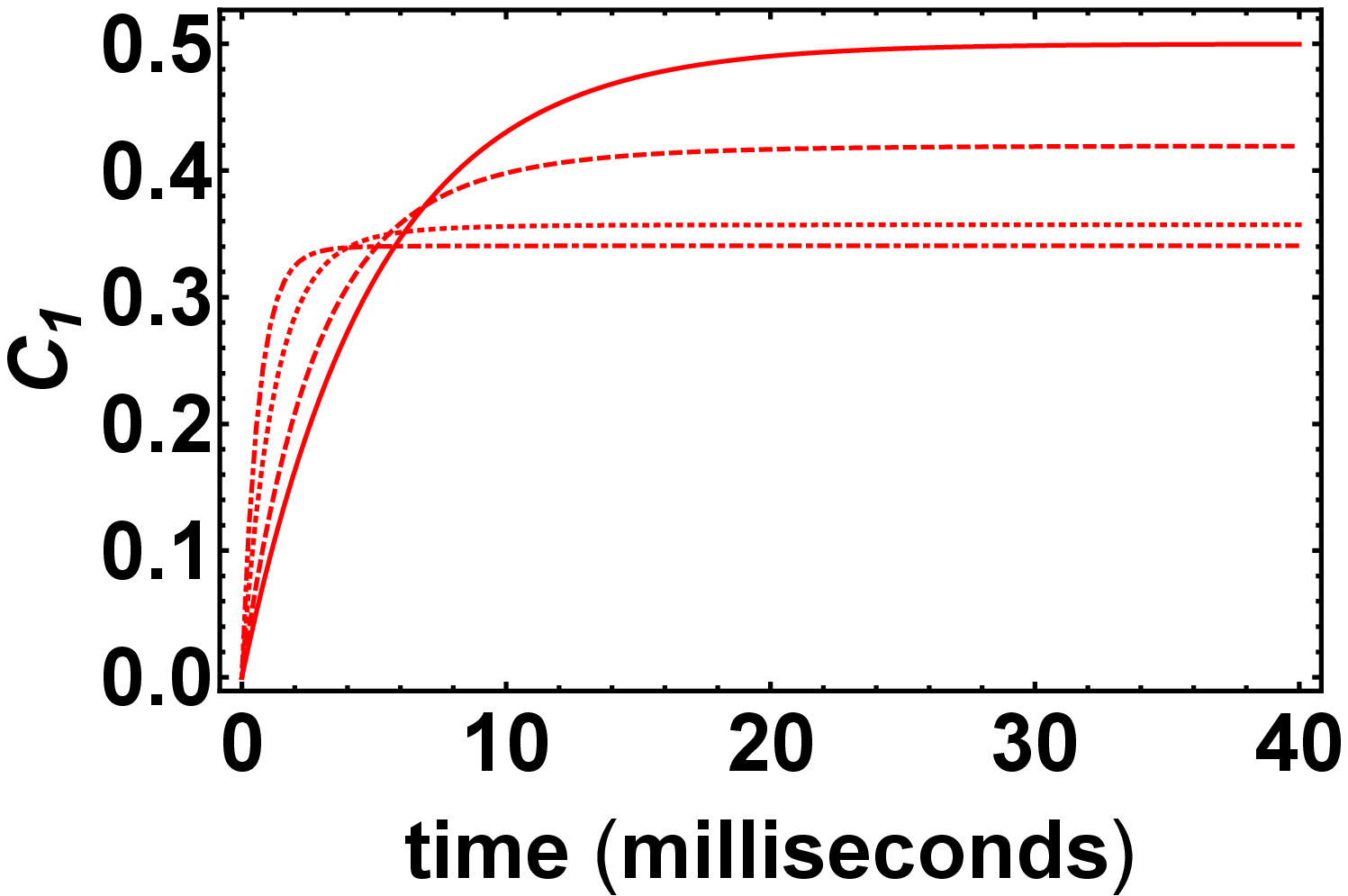}
		\label{fig:CohVsEps}
	}
	\subfloat[][]{%
		\includegraphics[width=0.23\textwidth]{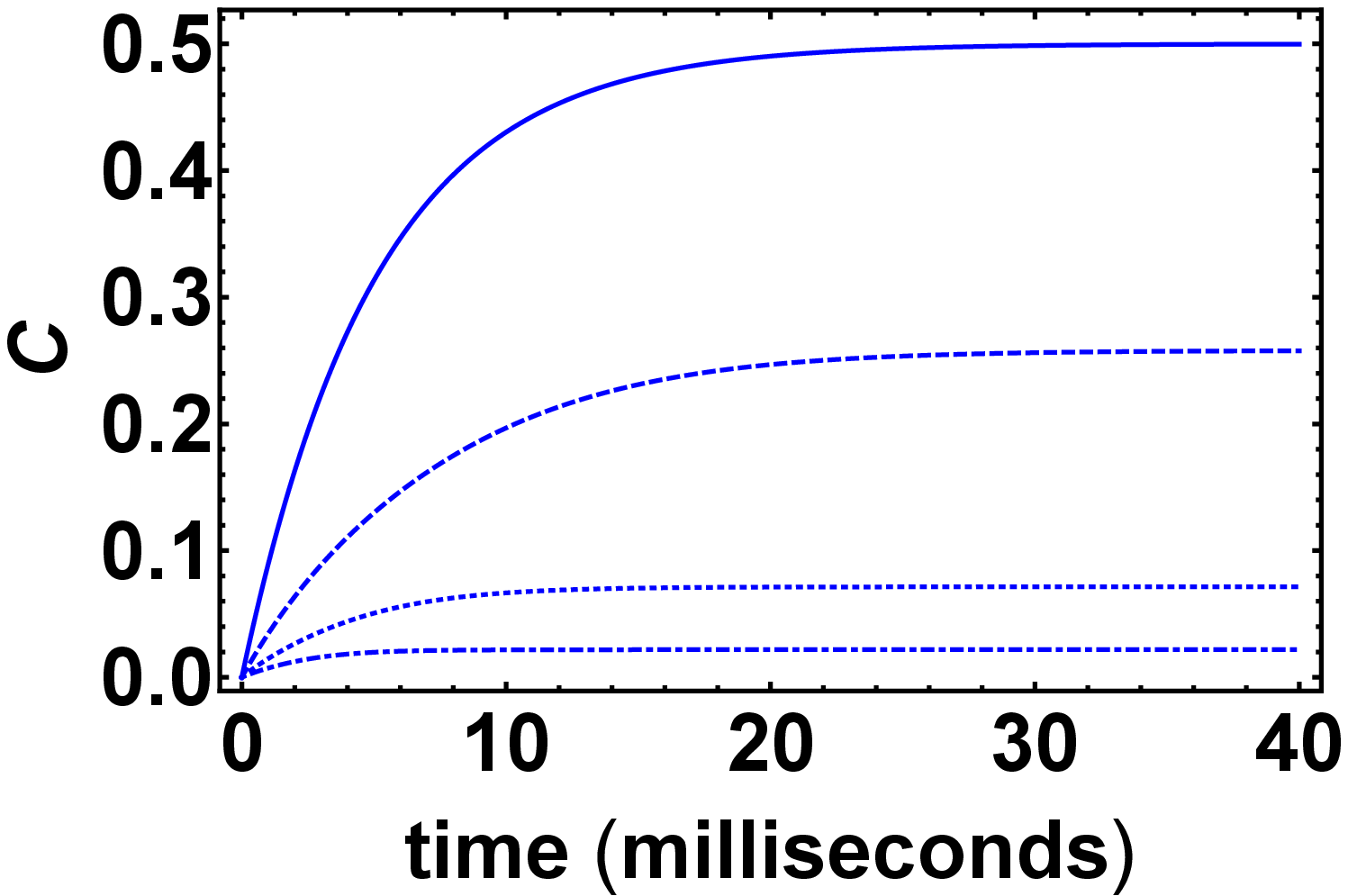}
		\label{fig:ConcVsEps}
	}
	\caption{\label{fig:CorrVsEps} (Color online) Dynamics of the $l_1$-norm coherence $C_1$ (red curves in (a)) and Concurrence $C$ (blue curves in (b)) of two NV center qubits in a quasi-one-dimensional thermal magnon bath with injected coherence $\epsilon=0$ (solid red and blue curves), $\epsilon=0.5$ (dashed red and blue curves),  $\epsilon=1.5$ (dotted red and blue curves), and $\epsilon=1$  (dotdashed red and blue curves). The other parameters are $T=1$ mK, $E=0$, $z_{\text{NV}}=20\,$ {nm}, $L_x=1.236\,\mu$m, $ L_z=20$ nm $T_1,T_2^\ast\rightarrow\infty$ s, 
	$x_{1,2}=\pm L_x/4$ m.}
\end{figure}

Influence of the coherence parameter, $\epsilon$ on the entanglement and coherence dynamics is plotted in Fig.~\ref{fig:CorrVsEps}.  Fig.~\ref{fig:CorrVsEps} shows that both SSE and SSC decrease with the $\epsilon$. Steady-state is reached earlier at higher $\epsilon$. The rate to get the steady state is faster (slower) for SSC (SSE). While SSE gets arbitrarily small and vanishes at large $\epsilon$, SSC saturates to $\sim 0.33$, same as the saturation value at high temperatures. The decrease in SSE and SSC is inevitable. Effective temperature character of $\epsilon$ populates the excited state, and hence the occupations of the $\ket{eg},\ket{ge}$ levels decrease, limiting the possible quantum coherence between these degenerate levels. The surviving coherent steady-state is however not an entangled state. In Fig.~\ref{fig:CorrVsEps}, we present the range of $\epsilon$ beyond
the physically feasible values of $0< \epsilon <0.7$ to show the general
behavior more clearly. The physical range of $\epsilon$ is restricted by the $B_1$ dependence of $\epsilon$. The larger $\epsilon$ values demand
the larger $B_1$, which is restricted by the saturation field value of $\sim 0.5$ T beyond which the YIG is demagnetized. With the calculated $s$ and $B_0$ values, and taking $N=10^3$, we find maximum $\epsilon\sim 0.7$.   

The effect of temperature on the entanglement and coherence dynamics is the same as that of  $\epsilon$, hence it is not shown here. We only remark that within the whole temperature range, of $0$ K to $0.5$ K, limited by the two-level NV qubit assumption, significant SSC can be obtained, while SSE requires much lower ($\sim 1-10$ mK) temperatures. In Sec.~\ref{sec:model-spinChainMagnons}, we assumed that NV center, whose ground state
is a spin-triplet $\ket{S=1,m_S=0,\pm 1} \equiv \ket{m_S}$, 
can be described as a qubit of $\ket{0}$ and $\ket{-1}$ states. To restrict the dynamics of the NV center to the manifold of qubit states, we require that
the state $\ket{+1}$ will always have negligible population, which can be ensured by
using sufficiently low temperatures and a bias magnetic field to separate the energy levels. The energy of the state $\ket{+1}$ is $\hbar(D+\gamma_{\mathrm{NV}}B_0)$. Transitions to the $\ket{+1}$ state from the $\ket{0}$ state can be neglected if there are negligible number of magnons with the sufficient energy, which is $\hbar(D+\gamma_{\mathrm{NV}}B_0)$. Using Bose-Einstein distribution for the mean
number of magnons $\bar{n}$ and taking $B_0\sim 51$ mT, we find the operating temperature as $T< 0.5$ K to satisfy $\bar{n}<0.1$.
At higher temperatures, the mean number of magnons resonant with the (dressed) qubit and the $\ket{0}$-$\ket{+1}$ transitions becomes comparable. While the operation temperature for the two-level NV center assumption can be as high as  $T< 0.5$ K, that does not mean we can get entanglement at such high temperatures.

For the given initial condition, when there are no private baths, the time dependent state is always of the form
\begin{eqnarray}\label{eq:rho_NoLocalNVbaths}
\rho(t)=\left(
\begin{array}{cccc}
	 \rho_{11}(t) & 0 & 0 & 0 \\
	0 & \rho_{22}(t) & \rho_{23}(t) & 0 \\
	0 & \rho_{32}(t) & \rho_{33}(t) & 0 \\
	0 & 0 & 0 & \rho_{44}(t) \\
\end{array}
\right),
\end{eqnarray}
where the elements of $\rho(t)$ are indicated by $\rho_{ij}(t)$ with $i,j=1..4$.
We use the standard basis $\{\ket{1}\equiv\ket{11},\ket{2}\equiv\ket{10},\ket{3}\equiv\ket{01},\ket{4}\equiv\ket{00}\}$ with $\ket{+}\equiv\ket{1}$ and $\ket{-}\equiv\ket{0}$. $\rho_{22}>\rho_{33}$ for $\rho(0)=\ket{10}\bra{10}$ and $\rho_{22}<\rho_{33}$ for $\rho(0)= \ket{01}\bra{01}$. 
The elements are always real so that $\rho_{23}(t)=\rho_{32}(t)$  and we found that $\rho_{23}(t)<0$. At low temperatures ($T\le 10$  mK), the elements tend to $\rho_{11}=0$, $\rho_{44}=0.5$ and $\rho_{ij}=0.25$ with $i,j\in\{2,3\}$ at the steady-state, for which $C_1=C=0.5$. 

For the state in Eq.~(\ref{eq:rho_NoLocalNVbaths}) we have $C_{_1}=2|\rho_{23}(t)|$, approaching to $0.5$ in the steady state. Accessibility and generation of only $\rho_{23}$ and not the other coherences by thermal means is not surprising from the point of view of the classification of coherences with respect to their thermodynamic heat and work equivalents~\cite{Dag2016,Tuncer2020,Latune2019,Latune2019b,Latune2020}. 
Coherence $\rho_{23}$ belong to the class of so-called heat-exchange coherences~\cite{Dag2016,Tuncer2020}. Considering their resource value for quantum information engines, steady state generation of these coherences makes our scheme significant for quantum information thermodynamics applications, too. 

\subsubsection{Decoherence free subspaces of NV center qubits}\label{sec:DFS_SSE_noNVlocalChannels}

To appreciate the significance of the structure and the long time robustness of $\rho(t)$, let's determine the states spanning the DFS of the qubits-magnon bath overall system. For that aim we determine the eigenvectors of
the system operator in Eq.~(\ref{eq:HsbCompact}). For symmetric placement of the qubits
about the center of the chain we can drop the qubit index $i$ from the bath operators and write Eq.~(\ref{eq:HsbCompact}) as 
\begin{eqnarray}\label{eq:HsbCompact2}
\hat{H}_{\text{SB}}(t)=
\sum_{\alpha}\hat{S}^\alpha(t)\hat{B}^\alpha(t),
\end{eqnarray}
in terms of the collective spin operators
\begin{eqnarray}
\hat{S}^\alpha(t)=\sum_{i}\hat\sigma_i^\alpha(t).
\end{eqnarray}
Besides, when we plot the interaction coefficients $\xi_k,\eta_k,\zeta_k$ with respect to $k$, for the placement of qubits away from the ends of the chain, we see in
Figs.~\ref{fig:etak}-\ref{fig:xik} that
they are approximately real valued for the long wavelength modes ($k\sim 0$).
Moreover, we have the relations $\xi(k)\approx 0$, and $\eta(k)=-\zeta(k)$ 
for $k\sim 0$. Hence, using the Eq.~(\ref{eq:bathOps}), we find $B^z=0$ and $B^+=-B^-$,
which gives 
\begin{eqnarray}
\hat{H}_{\text{SB}}(t)\approx (\hat{S}^+(t)-\hat{S}^-(t))\hat{B}^+(t),
\end{eqnarray}
for $k\sim 0$. 

\begin{figure}[t!]
	\centering
	\subfloat[][]{%
		\includegraphics[width=0.48\linewidth]{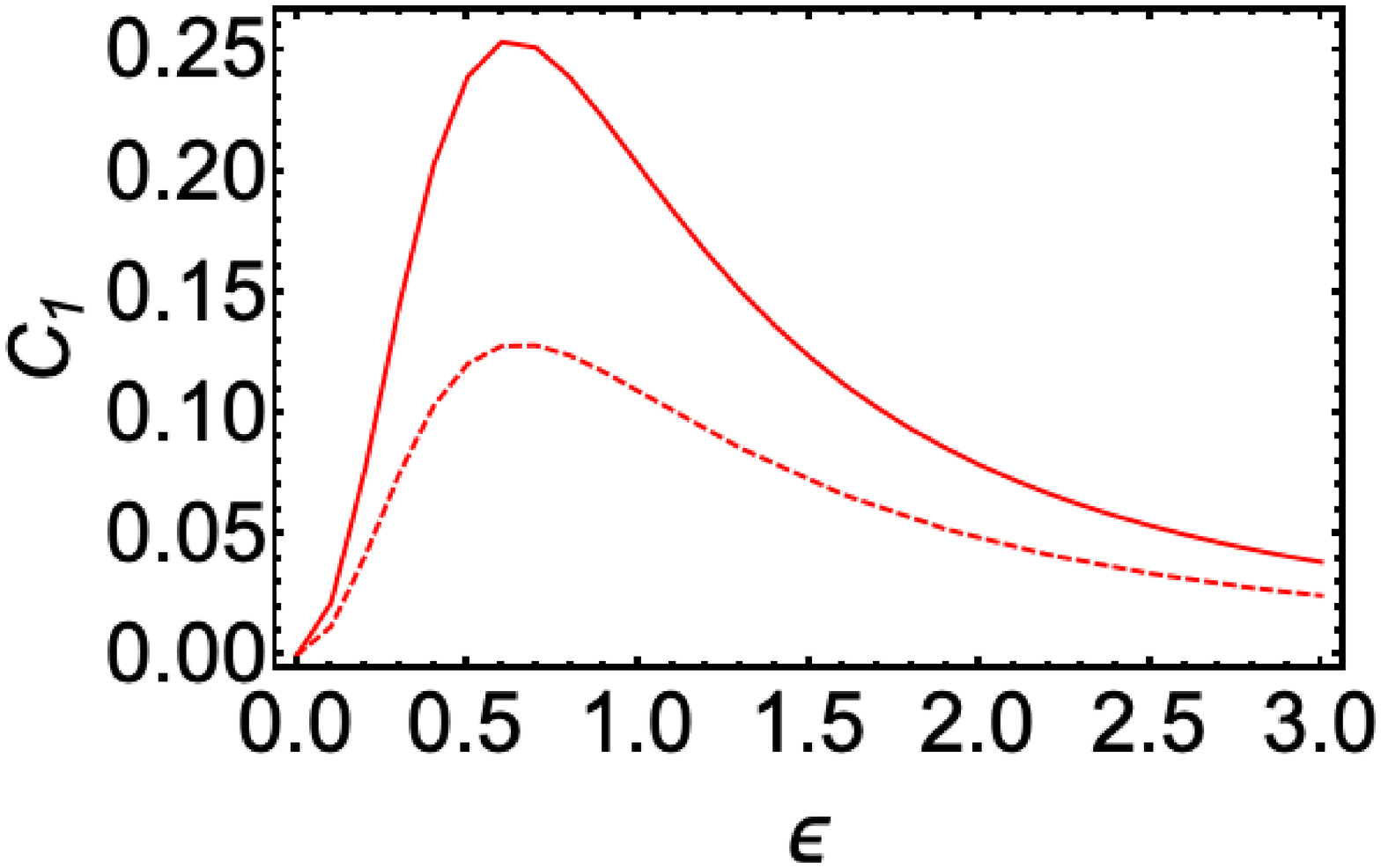}
		\label{fig:C1VsEps}
	}
	\subfloat[][]{%
		\includegraphics[width=0.50\linewidth]{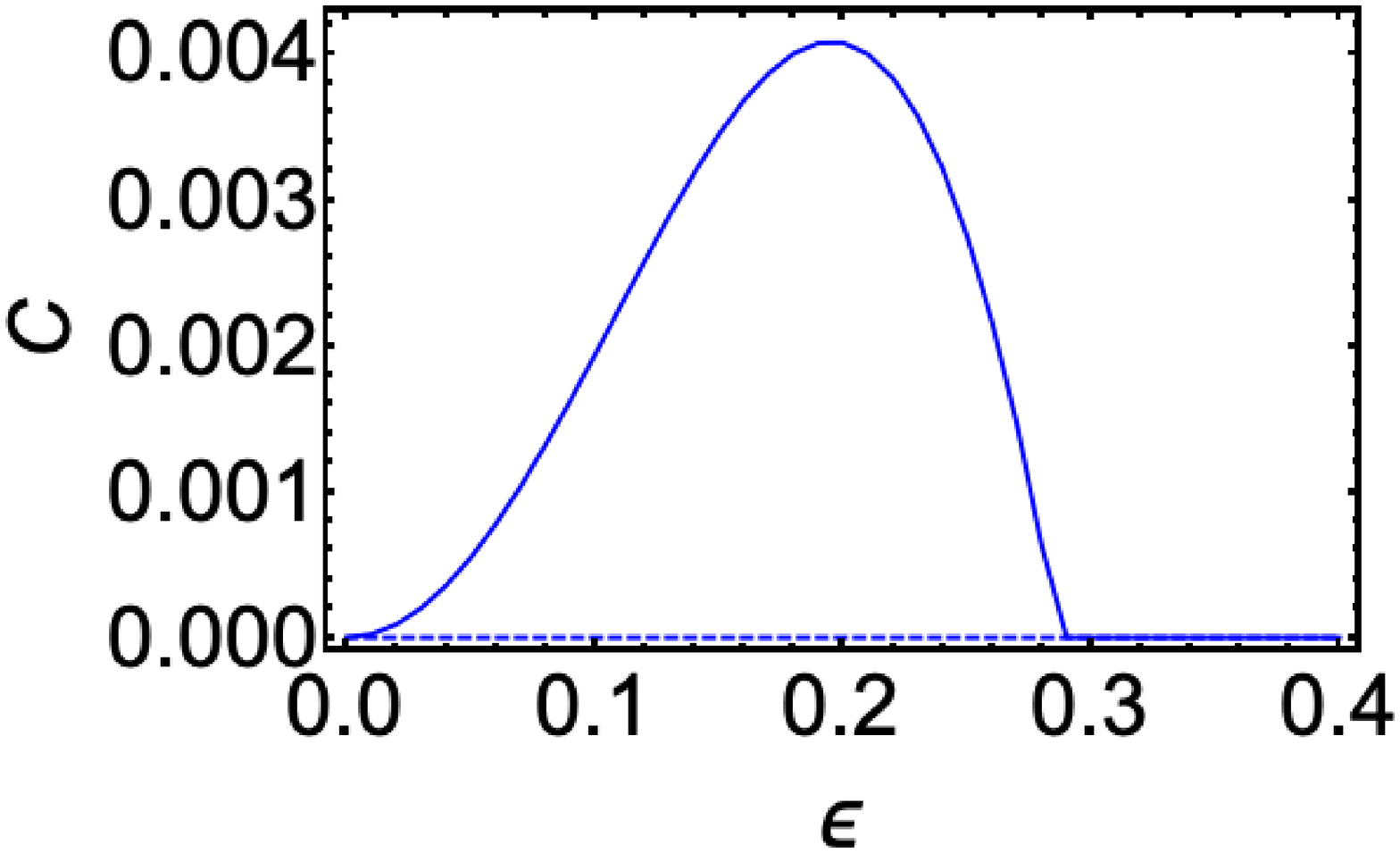}
		\label{fig:CVsEps}
	}
	\caption{\label{fig:C1CVsEps} (Color online) 
		Steady state behavior of the (a) $l_1$-norm coherence $C_1$ and (b) Concurrence $C$ of two NV center qubits in a public quasi-one-dimensional thermal magnon bath, with injected coherence $\epsilon$, when there are either (solid red and blue curves, $T_1=1$ ms) dissipative or dephasing (dashed red and blue curves, $T_2^\ast=1$ ms) private baths of the qubits. The dashed blue curve in (b) is a flat line at $0$. The other parameters are $L_z=20$ nm, $L_x=1.236\,\mu$m, $z_{\text{NV}}=5$ nm, $T=1$ mK, $x_{1,2}=\pm L_x/4$ m, $E=0.157241$ V/nm.}
\end{figure}

We can find the eigenvectors of the system operator $\hat{S}^+(t)-\hat{S}^-(t)$ to determine the DFS.
In terms of the collective spin states, one member of the DFS 
is the spin singlet state (we denote it by $\ket{\mathrm{DFS}_1}$),
\begin{eqnarray}\label{eq:DFS1}
\ket{\text{DFS}_1}=\ket{S=0,m_s=0}=\frac{1}{\sqrt{2}}(\ket{+-}-\ket{-+}).
\end{eqnarray}
This is the unique state that will be in the DFS for all $k$, while the spin triplet states cannot be in DFS in general, as they are not eigenvectors of the all the system operators $S^\alpha$. In our scheme, dynamics is restricted
over the $k\sim 0$, and hence an additional state, denoted by $\ket{\mathrm{DFS}_2}$ 
is added to the DFS,
\begin{eqnarray}\label{eq:DFS2}
\ket{\text{DFS}_2}&=&\frac{\ket{S=1,m_s=1}-\ket{S=1,m_s=-1}}{\sqrt{2}}\\
&=&\frac{1}{\sqrt{2}}(\ket{++}-\ket{--}).
\end{eqnarray}

We conclude that the evolution of the initial state $\ket{+-}$ yields states
$\rho(t)$ in the form in Eq.~(\ref{eq:rho_NoLocalNVbaths}) which is a
mixture of $\ket{\mathrm{DFS}_1}$ and $\ket{--}$ at all times, with relatively much smaller contribution from $\ket{++}$. Spin singlet is also the eigenstate of the free Hamiltonian
of the system with zero eigenvalue, hence both the dissipators and the free Liouvillian of the open system cannot change the dynamics out of the manifold of
the  $\ket{\mathrm{DFS}_1}$ and $\ket{--}$.  The fraction
of the DFS state grows in time and SSE is established. We remark that if the 
initial state is $\ket{\mathrm{DFS}_1}$ then it is always protected with $C(t)=1$.
Other entangled states, such as symmetric Bell state, would decay.  

Though $\ket{\mathrm{DFS}_2}$ has no effect on the SSE generated for the initial state $\ket{+-}$ when there is only the public magnon bath, it plays the decisive role to protect SSE against additional
decoherence channels from other private (nuclear spin) baths of the qubits. 

From quantum thermodynamical point of view, the coherences in $\ket{\mathrm{DFS}_2}$
are classified as work-like coherences or squeezing-type coherences. They are not
accessible by only thermal means. When we introduce $\boldsymbol{B_1}$ and
inject coherence into the bath, the squeezing-like dissipators
can induce dynamics to access
these elements (cf. the first two dissipators in Eq.~(\ref{eq:masterEqFinal})) to bring additional protection via  $\ket{\mathrm{DFS}_2}$, as we point out in the next section.

\subsubsection{NV center qubits in a public magnon and private nuclear spin baths}

\begin{figure}[t!]
	\centering
	\subfloat[][]{%
		\includegraphics[width=0.23\textwidth]{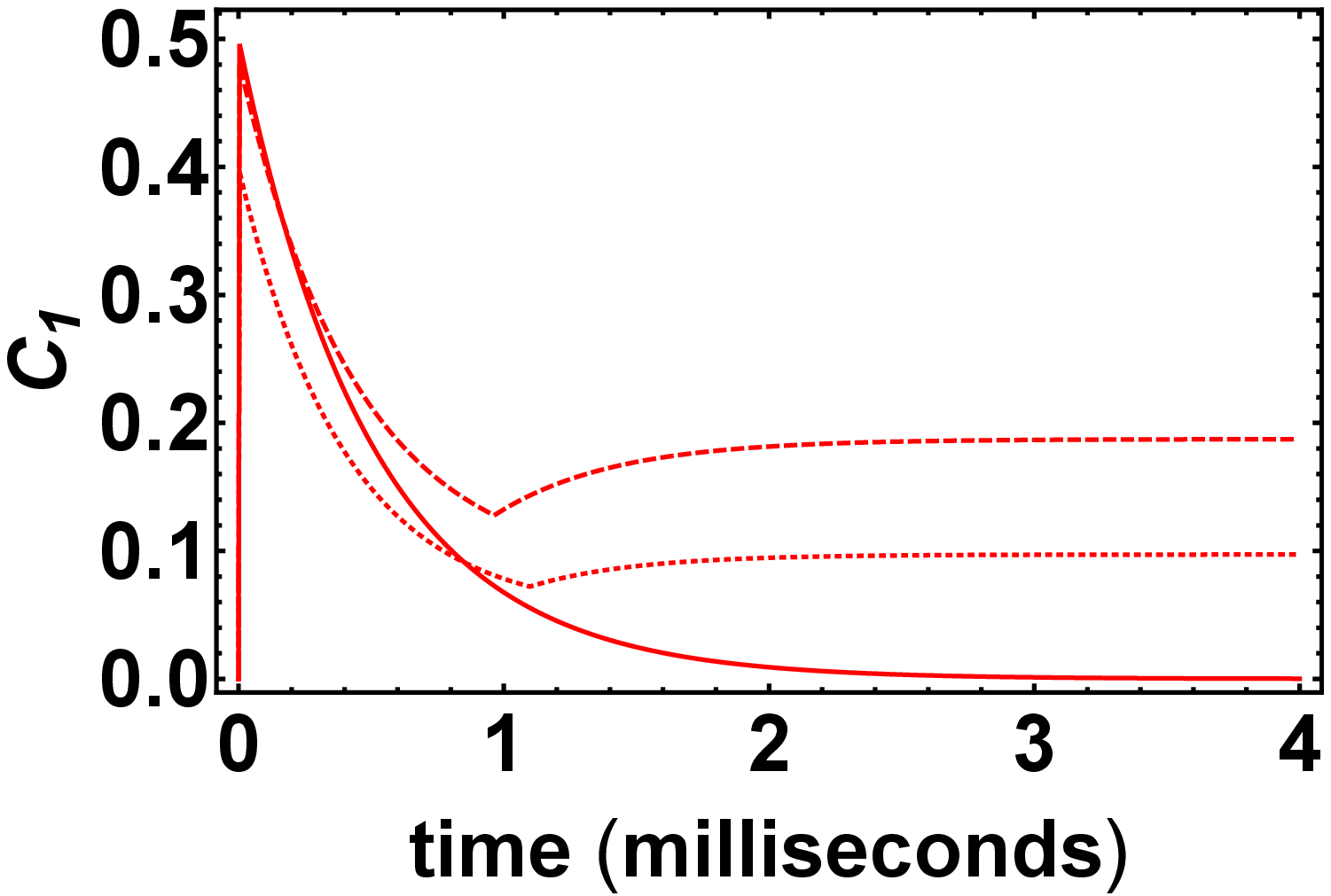}
		\label{fig:CohLowPrecE}
	}
	\subfloat[][]{%
		\includegraphics[width=0.23\textwidth]{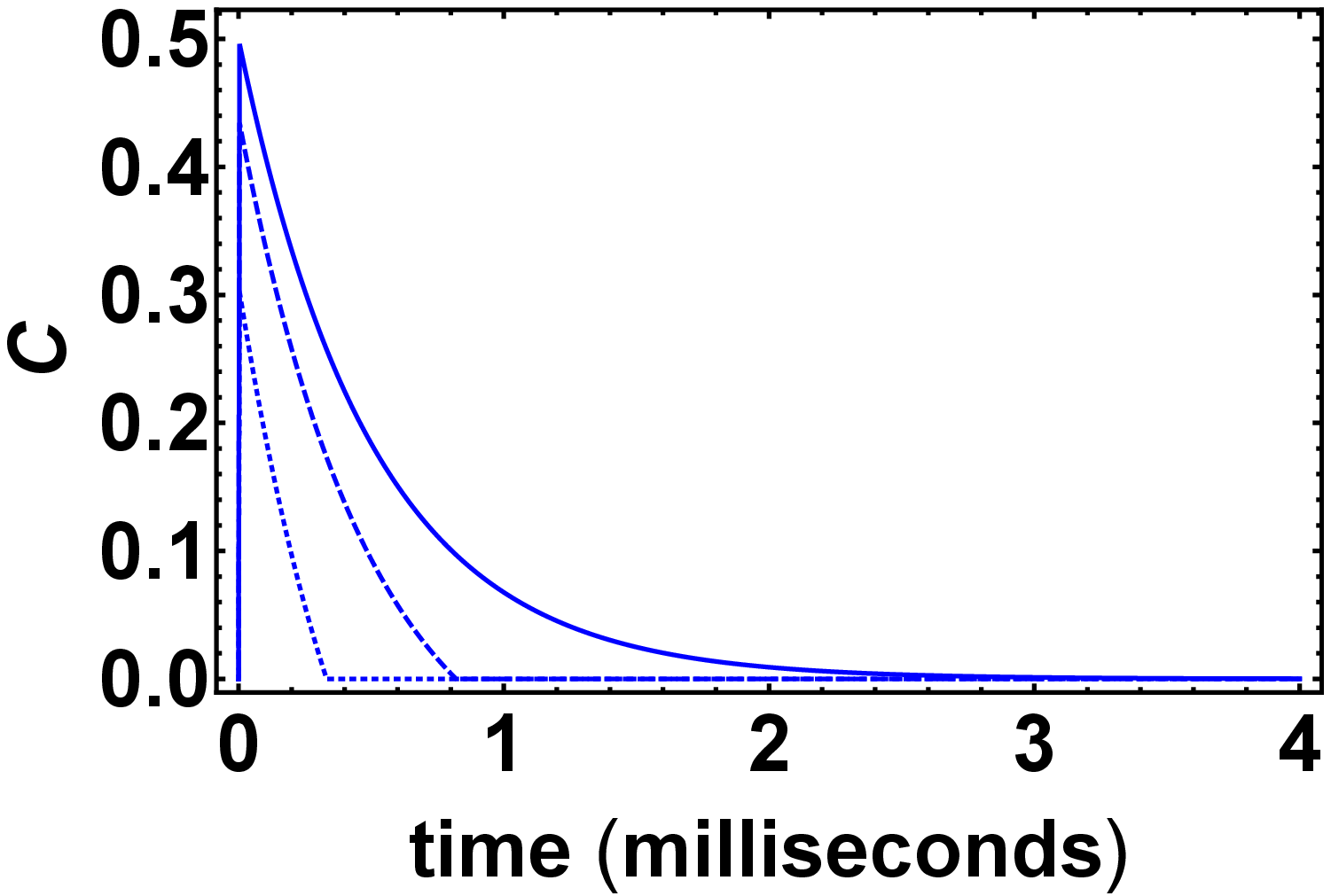}
		\label{fig:ConLowPrecE}
	}
	\caption{\label{fig:CorrLowPrecE} (Color online) 
		Dynamics of the (a) $l_1$-norm coherence $C_1$ and (b) Concurrence $C$ of two NV center qubits in a quasi-one-dimensional thermal magnon bath, with injected coherence $\epsilon=0$ (solid red and blue curves), $\epsilon=0.6$ (dashed red curve in panel (a)), $\epsilon=0.2$ (dashed blue curve in panel (b)), $\epsilon=1.5$ (dotted red curve in panel (a)), and $\epsilon=0.4$ (dotted blue curve in panel (b)). The other parameters are  $z_{\text{NV}}=5$ nm $L_z=20$ nm, $L_x=1.236\,\mu$m, $T=1$ mK, $T_1,T_2^\ast=1$ ms, $x_{1,2}=\pm L_x/4$ m, $E=0.157241$ V/nm.}
\end{figure}

Behavior of SSC and SSE with the injected coherence is plotted in Fig.~\ref{fig:C1CVsEps}. Coherence of the magnon bath has two competing effects on the dynamics of qubit-qubit correlations. First, bath coherence can effectively increase the bath temperature perceived by the qubit system and hence decrease the quantum correlations. Second, bath coherence can produce the effective drive and squeezing effects on the qubits. Simultaneous existence of the positive and negative influences of the bath coherence suggests that we can expect that there can be critical coherence values for which SSE and SSC can be possible and optimal when there are private baths.
Fig.~\ref{fig:C1CVsEps} confirms that intuitive expectation. In contrast to the case of single public bath, presence of private baths yield a non-monotonic behavior of SSE and SSC with injected coherence to the public bath. We see that critical values of $\epsilon\sim 0.6$ and $\epsilon\sim 0.2$, are different, respectively, for SSC and SSE. Besides, the critical $\epsilon$ values are insensitive to the type of the decoherence channel. In addition, distribution of SSC values with $\epsilon$ is broader for SSC relative to SSE. SSE drops sharply to zero after the critical $\epsilon$ in contrast to the slow change of SSC towards a finite saturation value beyond its maximum.

\captionsetup{labelfont=bf,font=small}
\begin{figure}[ht!]
	\centering
	\subfloat[][]{%
		\includegraphics[width=0.50\linewidth]{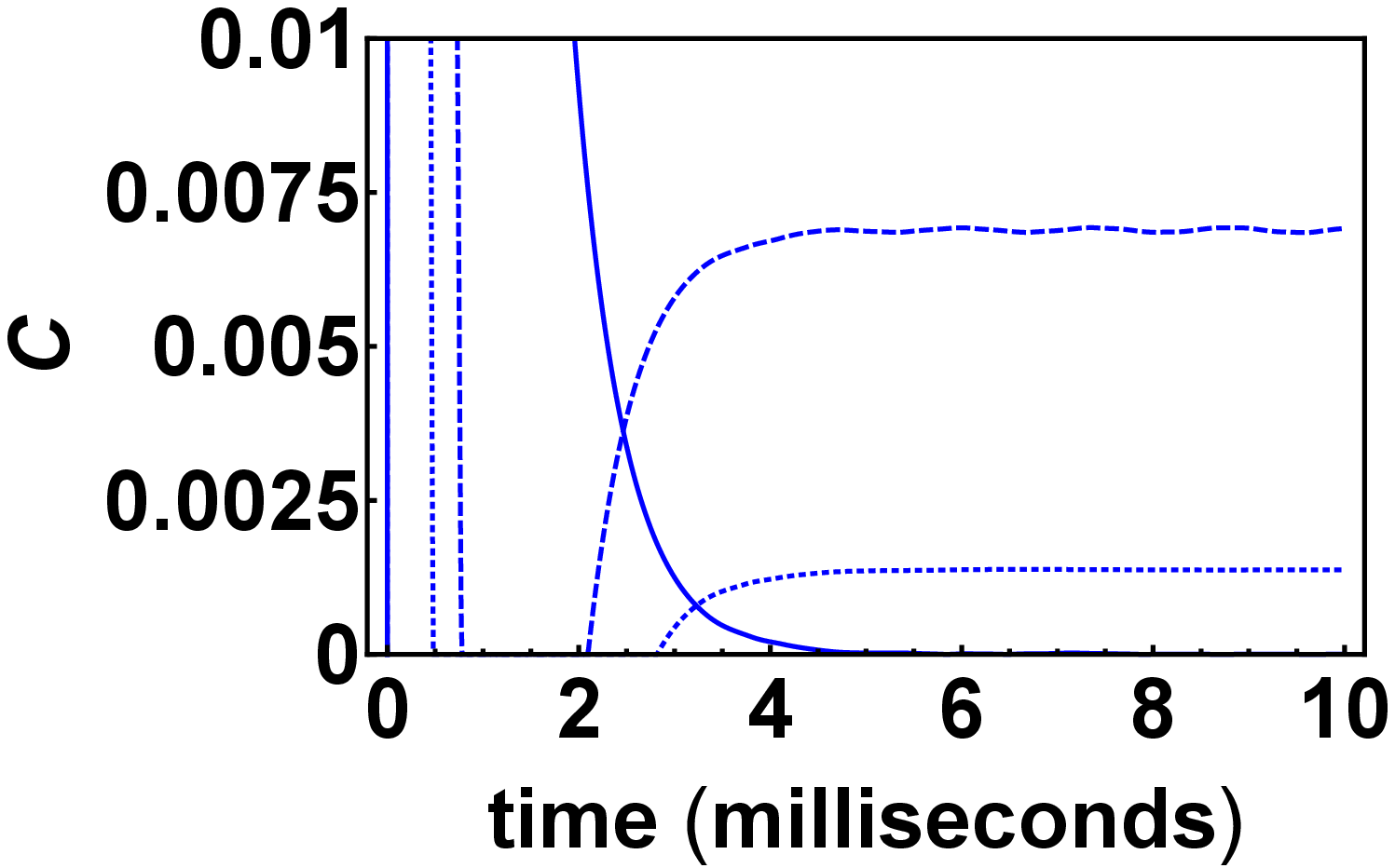}
		\label{fig:ChighPrecE}
	}
	\subfloat[][]{%
		\includegraphics[width=0.49\linewidth]{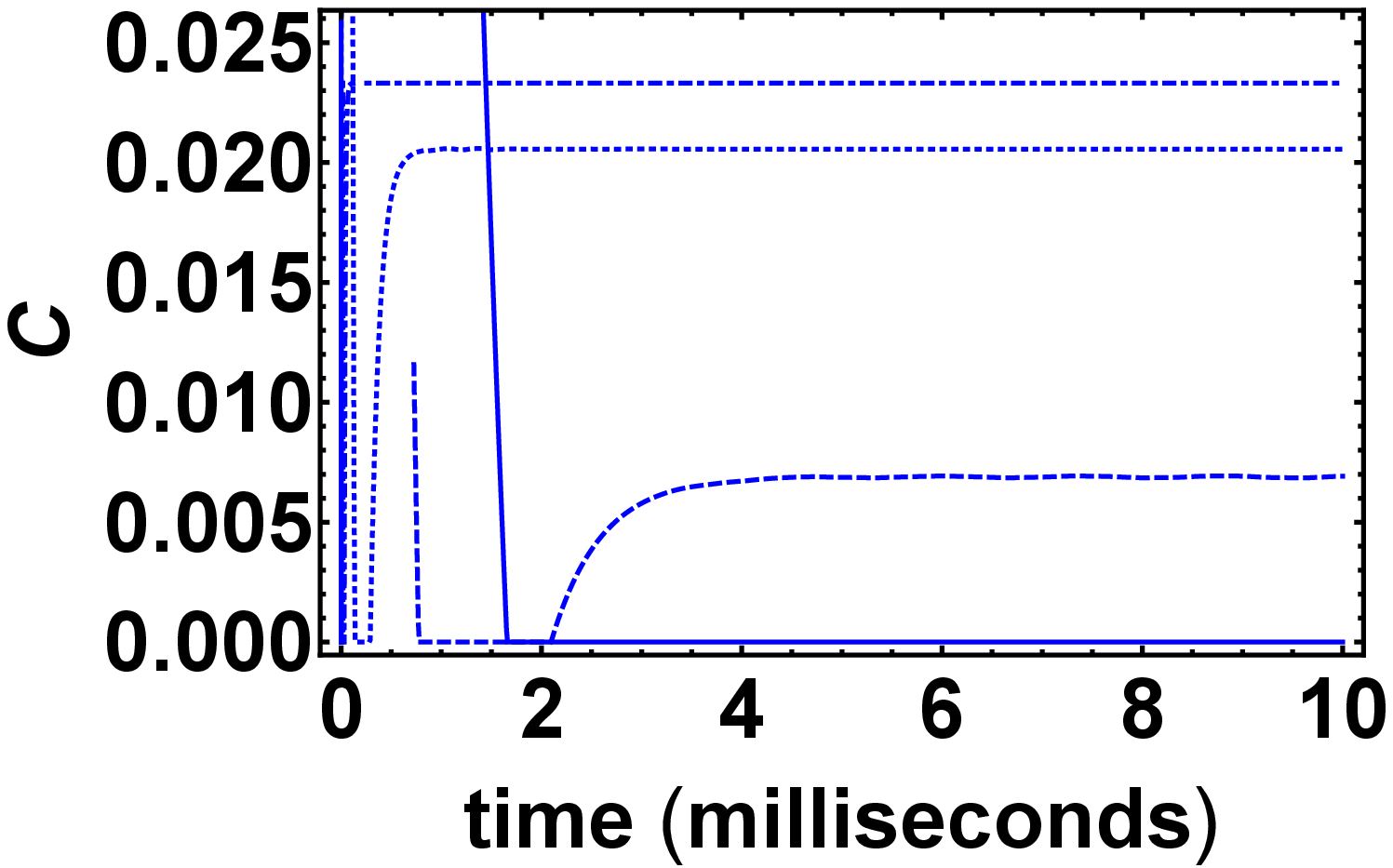}
		\label{fig:curiousT1}
	}
	\caption{ (Color online) 
		a) Concurrence $C$ of two NV center qubits in a quasi-one-dimensional thermal magnon bath with injected coherence $\epsilon=0$ (solid blue curve), $\epsilon=0.2$ (dashed blue curve),  and $\epsilon=0.3$ (dotted blue curve).The other parameters are  $z_{\mathrm{NV}}=5$ nm $L_z=20$ nm, $L_x=1.236\,\mu$m, $T=1$ mK, $T_1,T_2^\ast=1$ ms, $x_{1,2}=\pm L_x/4$ m, $E=0.1572412$ V/nm. b) Dynamics of the Concurrence $C$ of two NV center qubits in a quasi-one-dimensional thermal magnon bath with injected coherence $\epsilon=0.2$ for the qubits' longitudinal relaxation rates $T_1=1$ s (solid blue curve), $T_1=1$ ms (dashed blue curve), $T_1=1\,\mu$s  (dotted blue curve)  (dotdashed blue curve). The other parameters are  $z_{\mathrm{NV}}=5$ nm $L_z=20$ nm, $L_x=1.236\,\mu$m, $T=1$ mK, $T_2^\ast=1$ ms, $x_{1,2}=\pm L_x/4$ m, $E=0.1572412$ V/nm.}
\end{figure}

In Fig.~\ref{fig:C1CVsEps}, we analyze the role of dissipative and dephasing private baths separately. When the dissipative private channels are acting alone, both SSE and SSC can be obtained. The value of $E=0.157241$ V/nm is determined by considering the minimum precision required to make $L_z-L_e$ sufficiently low to increase the DOS, which is translated to the enhanced dissipation rate $\kappa$ that gives SSE. The idea of fine tuning external homogeneous magnetic field for sizable effective qubit-qubit coupling by eliminating the bath degrees of freedom with Schrieffer-Wolff transformation~\cite{Bravyi2011} has already been proposed~\cite{Trifunovic2013}. Our approach is similar but for the case of bath-mediated qubit-qubit coupling. In addition to  resonance tuning with magnetic field, we propose to control effective YIG film thickness via external electric field to get competitive dissipation rates of the public bath against the private decoherence channels. On the contrary, when the dephasing private baths act alone, SSE entanglement cannot be established for any $\kappa$, and the injected coherence has no positive effect. This cannot be improved by decreasing the YIG strip thickness effectively using the electric field. 

We plot the case of simultaneous presence of both private decoherence channels in Fig.~\ref{fig:CorrLowPrecE} for the same level of precision in $E=0.157241$ V/nm.
The conclusion of Fig.~\ref{fig:C1CVsEps} remains the same. SSC saturates to its optimal value at the critical $\epsilon\sim 0.6$ of SSC; while no SSE is obtained even for the critical $\epsilon\sim 0.2$ of the case of SSE with only private dissipations.

When both dephasing and dissipative private channels are open, if we increase the precision of tuning $L_z$ and $L_E$  with another digit using $E=0.1572412$ V/nm, we can obtain SSE, as shown in Fig.~\ref{fig:ChighPrecE}, at the critical $\epsilon\sim 0.2$ of the case of SSE with only private dissipations. This suggest that the critical $\epsilon$ values obtained when the private dissipation acts alone can be used when the private dephasing is also on. Lack of SSE when the private dephasing channels are acting alone, and emergence of SSE when both dissipative and dephasing channels are present can raise the curious question if increasing the private dissipation can give higher SSE. Fig.~\ref{fig:curiousT1} gives a positive answer to this question. Remarkably, this is a hypothetical case of academic interest as normally the longitudinal relaxation is slower than the transverse relaxation, though some engineering of $T_1$ may be possible using applied fields on NV centers, similar to those methods used for quantum dots~\cite{Amasha2008}. Promising developments in probing and engineering nuclear spin baths of NV centers should be noted, too~\cite{Jackson2021}. Nevertheless, Fig.~\ref{fig:curiousT1} reveals that there is a saturated maximum SSE with $C\sim 0.025$, when $T_1$ gets faster towards to $\mu$s regime while $T_2$ remains in the ms regime. This intriguing conclusion, as well as our previous statements can be physically explained in terms of the DFS structure of the qubit system.

The steady state our on-chip device generates due to public bath mediated coupling is approximately a mixture of superposition of the pairwise ground state with a Bell state, 
$\rho_{\mathrm{SS}}=\ket{\psi_{\mathrm{Bell}}}
\bra{\psi_{\mathrm{Bell}}}+\ket{gg}\bra{gg}$, when there are only dissipative private channels. It is explicitly written as
\begin{eqnarray}
\rho _{\text{SS}}=\left(
\begin{array}{cccc}
a & 0 & 0 & 0 \\
0 & b & x & 0 \\
0 & x & c & 0 \\
0 & 0 & 0 & d \\
\end{array}
\right),
\end{eqnarray}
where $a\sim 0,b\sim c,d\sim 1$ and $x\in{\cal R}$.
Such a state has only single coherence, between the degenerate single qubit excitation states (also known as heat-exchange coherences~\cite{Dag2016,Tuncer2020,Latune2019,Latune2019b,Latune2020}). Protection of this coherence is provided by $\ket{\mathrm{DFS}_1}$ of Eq.~(\ref{eq:DFS1}). When the thermal magnon bath has injected coherence via the inhomogeneous magnetic field $\boldsymbol{B_1}$, we get $\rho_{\mathrm{SS}}\equiv\rho _{\mathrm{X}}$,
\begin{eqnarray}
\rho _{\text{X}}=\left(
\begin{array}{cccc}
a & 0 & 0 & y \\
0 & b & x & 0 \\
0 & x & b & 0 \\
y^\ast & 0 & 0 & d \\
\end{array}
\right),
\end{eqnarray} 
where we see that additional protection comes from $\ket{\mathrm{DFS}_2}$
of Eq.~(\ref{eq:DFS2}). The new coherence $y$
can only emerge when the squeezing-like dissipators of the 
master equation~(\ref{eq:masterEqFinal}). Without $y$, there is no SSE in the presence of private baths. It is therefore crucial to go beyond the standard form of the master equations for the weakly-coherent baths~\cite{Rodrigues2019}, and to keep
the second order terms in $\epsilon$ even if it is weak relative to the first
order effective drive term in the open system dynamics to properly assess the SSE and SSC. 

\section{Conclusion}\label{sec:conc}
We investigated steady-state entanglement and coherence generation between two NV center qubits using a common magnon bath in a YIG nanostrip static external fields
and its protection against local dehasing and dissipation channels. Our idea is to use beneficial effects of public bath to mediate entanglement between qubits against decoherence effect of private baths. To help the shared bath for this task, we discussed the bath dispersion and coherence engineering together with the role of system geometry, which can be compared to exploitation of capacitor geometry to increase its capacitance. 

Specifically we consider two NV center qubits on a YIG nanostrip as our example system. One external magnetic field is used to tune the magnetostatic mode of the YIG magnons to the qubit resonance while another magnetic field, transverse to the first one, is used to inject coherence into the thermal magnon bath. Magnitude and spatial profile of the coherence injecting field contributes to control the Markovian character of the open system dynamics. Additional electric field is used to effectively decrease the thickness of the YIG strip, allowing the tuning group velocity and the DOS at the magnetostatic mode, in return contributes to the sizeable magnon-mediated qubit-qubit interaction. We develop a generalized quantum master equation for our open system for weak coherences but keeping the coherence effects up to the second order, which brings squeezing-like dissipators next to the first order effective drive term. Such squeezing-like terms extend the decoherence free subspace of the qubits from Bell state singlet to a triplet, providing additional protection to the private dephasing and dissipation. We find a non-monotonic behavior of SSE and SSC with the injected coherence when private baths present so that critical coherences can be used to optimize the SSE and SSC. Curiously, the SSE increase when private longitudinal relaxation (dissipative decoherence channel) is present next to the private transverse (dephasing channel) relaxation. Dynamics of SSE and SSC are shown to be sudden death of correlations in the transient regime, followed by a delayed setting of quantum correlations in the steady-state. 

Detailed analysis of the interaction coefficients revealed that dephasing to the magnon bath is not effective when the qubits are placed away from the ends of the strip, and the interaction coefficients as well as the reported results remain the same uniformly so that our scheme can be generalized to multiple qubits placed on the strip in a straightforward manner. Further scaling to multi-qubit entanglement might be possible by using nanopatterned mesh of YIG strips~\cite{Liu2018} with NV qubits on top, though careful study of stray magnetic fields in addition to control fields is required to rigorously assess the extent of scalability. Tunable Markovian character of our scheme can allow for explorations of Markovian to non-Markovian dynamical regime transitions and effects of non-Markovianity on SSE and SSC. Furthermore, at the cost of energetic expenses, time-dependent fields and time-dependent master equations can be considered for increasing SSE and SSC. Effects of lateral dimension on the interaction coefficients are not included in our theory. Collective enhancement of interactions can be possible up to crically narrow ultrathin YIG strips, which can be another future study. 

In conclusion, we propose a hybrid magnonic device that can be tuned to operate as robust quantum coherence and entanglement generator between distant qubits in steady-state. Depending on technological progress to engineer magnon dispersion in ultrathin magnetic strips using external static electric and magnetic fields, our scheme can be promising for scalable coherence and entanglement generation and long-time protection for versatile quantum technology applications.

\section*{Acknowledgements}

The authors acknowledge support from TUBITAK Grant No.~$120$F$230$. M.~C.~O.~acknowledges support from TUBITAK Grant No.~$117F416$, TUBA-GEBIP Award from Turkish National Academy of Sciences (TUBA), and funding from the European Research Council (ERC) under the European Union’s Horizon 2020 research and innovation programme with grant agreement No.~$948063$ and project acronym SKYNOLIMIT.
\section*{Data Availability Statement}
All data generated or analysed during this study are included in this paper.

\appendix

\section{Magnons in a linear spin chain}\label{sec:appendix-magnons}

Here we present a short review of some fundamentals of magnons in a linear spin chain,
which can illuminate the differences and size effects in the dispersion relation
of the magnons in a YIG nanostrip. 

Magnons are quanta of collective spin excitations described as spin waves~\cite{Bloch1930,Holstein1940,Dyson1956}. Let's 
consider a linear chain of $N$ spins (we assume $N$ is even) 
modeled by the 
Heisenberg Hamiltonian
\begin{eqnarray}\label{eq:Heisenberg}
\hat H_\text{chain}=-\hbar\gamma_0B_0\sum_{j=-N/2}^{N/2}\hat S_j^z
-2\hbar J
\sum_{j=-N/2}^{N/2-1}\hat{\bm{S}}_j\cdot\hat{\bm{S}}_{j+1}\nonumber\\
\end{eqnarray}
where $\hbar J>0$ is the exchange integral determining the ferromagnetic coupling
of a spin at site $j=-N/2..N/2$ to its neighboring spins at a lattice constant 
distance $a$ (see Fig.~\ref{fig:fig1-ModelSystem}). Spin locations are given by
\begin{eqnarray}
x_j=[j-\mathrm{sign}(j)\frac{1}{2}]a,
\end{eqnarray}
with the sign function, $\mathrm{sign}(x)=+1,0,-1$ for $x>0,x=0,x<0$, respectively.
Spin angular momentum operator $\hat{\bf{S}}_j$ 
is taken dimensionless.
The spins are subject to a uniform, static, external magnetic field of magnitude $B_0$
aligned in $z$ direction. The first term in the model Hamiltonian is the Zeeman energy,
where $\gamma_0=g\mu_B/\hbar$ is the gyromagnetic ratio (in units of rad/Ts) defined in terms of the $g$-factor and 
the Bohr magneton $\mu_B$.

Using the Holstein–Primakoff transformation~\cite{Holstein1940}, and taking its weak excitation approximation, we have
\begin{eqnarray}
\hat S^+_j&\approx&\sqrt{2s}\hat m_j,
\quad
\hat S^-_j\approx\sqrt{2s}\hat m_j^\dagger,\label{eq:HolsteinPrimakoff1}\\
\hat S_j^z&=&s- \hat n_j,\label{eq:HolsteinPrimakoff2}
\end{eqnarray}
where $s$ is the total spin, same for all sites, and $\hat m_j$ ($\hat m_j^\dagger$) is the annihilation (creation) operator of a magnon quasiparticle at site $j$.
The number operator of the magnons at site $j$ is denoted by 
$\hat n_j:=\hat m^\dagger_j \hat m_j$. Low excitation condition, 
$n_j:=\langle \hat n_j\rangle\ll 2s$ is well satisfied at low temperatures and for large $s$ values. 

Fourier transformed magnon operators are given by
\begin{eqnarray}
\hat m_k&=&\frac{1}{\sqrt{N}}\sum_{j=-N/2}^{N/2} \text{e}^{-ik x_j}\hat m_j,
\end{eqnarray}
and their commutators obey the bosonic algebra. 
The Hamiltonian $\hat H_\text{chain}$ in the magnon
representation takes the form 
\begin{eqnarray}\label{eq:H_chain_kSpace}
\hat H_{\text{mag},0}=\hbar\sum_{k=-\infty}^{\infty} \omega_k \hat m^\dagger_k \hat m_k,
\end{eqnarray}
where the magnon dispersion relation is two-fold degenerate for $\pm k$ and it is given by
\begin{eqnarray}\label{eq:MagnonDispersion}
\omega_{k}=\omega_0+4Js(1-\cos ka),
\end{eqnarray}
where we dropped a constant $E_0=-4NJs^2$, and $\omega_0:=\gamma_0B_0$ is the angular frequency of the $k=0$ mode. 
Physically, magnon quasiparticles are associated with small transverse spin fluctuations behaving as a wave with such a dispersion relation. In the main text we use a more sophisticated magnon dispersion for our ultrathin YIG stripes due to finite size effects (cf.~Eq.~(\ref{eq:magnonDispersionFiniteSize})). 

From the dispersion relation, we evaluate the magnon density of states (DOS) $D(\omega)$ using
$D(\omega)d\omega:=4(Ldk)$, where the factor of $4$ comes from two-fold 
polarization and two-fold spatial ($\pm k$) degeneracies. 
We change the units of DOS to seconds for convenience, by including 
$L=(N-1)a$ in its expression, and write 
\begin{eqnarray}\label{eq:MagnonDOS}
D(\omega)=\frac{4}{a}\frac{1}{\sqrt{\omega-\omega_0}\sqrt{8Js-\omega+\omega_0}}.
\end{eqnarray}
Consistent with the low temperature assumption, significant modes can be taken those within the long wavelength limit $ka \ll 1$, for which the dispersion relation (\ref{eq:MagnonDispersion}) reduces to
$\omega_{k}=\omega_0+2Jsa^2k^2$. The DOS (\ref{eq:MagnonDOS}) for 
$ka\ll 1$ approximates to 
\begin{eqnarray}\label{eq:DOSapprox}
D(\omega)=\frac{2N}{\sqrt{2Js}}\frac{1}{\sqrt{\omega-\omega_0}}.
\end{eqnarray}
Square-root singularity of the DOS is typical for a free particle in one-dimensions.
As DOS directly contributes to the dissipation rates of a system through the Fermi's Golden Rule, it is exploited to enhance radiative decay in isotropic photonic 
crystals with a one-dimensional phase space, too. 
Infinitely large scattering or dissipation rates can be related to
the the zero group velocity at the band edge so that the time delayed response
of the bath is classified to be higly  non-Markovian~\cite{DeVega2017,Roy2010,Wang2011,Wang2012,Yang2013,Wang2014,Woldeyohannes2015,Li2015,Wu2016,Shen2019,Dinc2019,Ma2020,Sinha2020,Sinha_2020}, though transition between Markovian and non-Markovian regimes can have non-monotonic dependence on finite system parameters in a general structured bath~\cite{Ma2014,DeVega2017}. However, a
one-dimensional spin chain is an idealization and
one can only have a quasi-one dimensional system in practical implementations. 
We discuss a modified dispersion relation
to take into account the lateral size effects when we specify a magnetic material to set the physical parameters for our spin chain  in Sec.~\ref{sec:physParameters},
and find a regime where the dynamics of our physical system can be restricted to Markovian regime yet still gets the benefits of the band edge.

In the continuum limit ($N\gg1$), Hamiltonian in 
Eq.~(\ref{eq:H_chain_kSpace}) can be written as
\begin{eqnarray}\label{eq:H_chain_wSpace}
\hat{H}_\mathrm{mag}=\hbar\int_{-\infty}^\infty\,\frac{d\omega}{2\pi} 
D(\omega) \omega  \hat m^\dagger(\omega) \hat m(\omega),
\end{eqnarray}
where the integral limits can be taken at $\pm\infty$ by assuming 
$D(\omega)=0$ outside the magnon frequency band of 
$[\omega_0,\omega_0+8Js]$. In the main text, we discuss how external electric and magnetic fields can be used to engineer the DOS to control dissipation of the qubits into the common magnon bath (cf.~Eq.~(\ref{eq:dosStripE})).

\section{Diamond NV centers}\label{sec:appendix-NVqubits}

NV center is an optically active color defect center, 
consisting of a substitutional nitrogen impurity and a nearest neighbor carbon vacancy in diamond lattice~\cite{Doherty2013}.
Typically, many NV centers are produced in a diamond host. Nevertheless, it is possible to isolate a single defect center for example in a few nanometer nanodiamond~\cite{Bradac2010}. We consider a setup (cf.~Fig.~\ref{fig:fig1-ModelSystem}) where a single NV center in a host nanodiamond can be placed on a spin chain.

From the Nitrogen, bulk donor, and the three dangling bonds of Carbon atoms around the Carbon vacancy, negatively charged NV center's electronic bound states consists of $6$ electrons and can be described as a spin-1 system~\cite{Doherty2013}.
NV center ground state is a spin triplet ($^3\mathrm{A}_2$) $\ket{Sm_S}$ with $S=1$ and
$m_S=0,\pm1$. The excited-state triplet ($^3\mathrm{E}$) is at $~1.95$ eV higher above $^3\mathrm{A}_2$~\cite{Weber2010} and will not be considered here. Accordingly, we write the single NV center Hamiltonian as 
\begin{eqnarray}\label{eq:Hnv}
\hat H_{\mathrm{NV}}=\hbar D \hat S_z^2 + \hbar \gamma_{\mathrm{NV}} B_0 \hat S_z,
\end{eqnarray}
where $D/2\pi=2.87$ GHz is the zero field splitting by the spin-spin interactions and $\gamma_{\mathrm{NV}}/2\pi=28.02$ GHz/T is the gyromagnetic ratio of the NV center with $g\approx 2$~\cite{Maze2011}, which is approximately the same as
$\gamma_0/2\pi=g\mu_B/2\pi\hbar=27.99$ GHz/T. While NV centers are subject to the $\boldsymbol{B_0}$, applied along the $z$-axis, we assume NV centers are
away from the range of influence of $\boldsymbol{B_1}$. This assumption is not a serious limitation in our theory as its effect would be an extra shift in the transition frequency of the qubits, which will be compensated by the resonance condition between the magnons and the NV center qubit. We take into account the shift in the qubit transition frequency due to the magnon field an neglect the shift by $\boldsymbol{B_1}$ for simplicity.
Spin-1 operators (dimensionless) are denoted by $S_\alpha$ with $\alpha=x,y,z$. In order to get the second (Zeeman) term of the Hamiltonian without $\hat S_x$ and $\hat S_y$, one of the molecular frame NV-axes must coincide with the lab frame $z$-axis~\cite{Muhlherr2019}. We assume nanodiamond crystal is oriented in such a way that the NV center's principal symmetry axis ($[111]$ crystal axis) is the same with the lab frame $z$-axis~\cite{Wu2018}.

\begin{table*}[t!]
	\begin{tabular}{ |p{4cm}|p{4cm}||p{4cm}|p{4cm}|}
		\hline
		\multicolumn{4}{|c|}{Parameter List} \\
		\hline
		$B_{0}$  				 					&  $51.16$ mT &
		$\gamma_{0}/2\pi\approx\gamma_{NV}/2\pi$    &  $28.02$ GHz / T \\
		\hline
		$\omega_{0}/2\pi$          					&  $1.4335$ GHz &	
		$\omega_{NV}/2\pi$							&  $1.4365$ GHz \\		
		\hline
		$\omega_{i}/2\pi$   						&  $1.4335$ GHz &
		$\Omega_{i}/2\pi$        					&  $1.4335$ GHz \\
		\hline
		$J/2\pi$   									&  $33.42$ GHz &
		$s$    										&  $14.2$ \\
		\hline
		$T$       									&  $0-0.5$ K &
		$D/2\pi$   									&  $2.87$ GHz \\
		\hline
		$L\equiv L_x$      							&  $1.24$ $\mu$m &
		$L_y$       								&  $120$ nm \\
		\hline
		$L_z$      									&  $20$ nm &
		$a$       									&  $12.376$ \AA \\
		\hline
		$N$   										&  $10^3$ &
		$z_{NV}$    								&  $5-20$ nm  \\
		\hline
		$x_1=L/4,x_2=-L/4$    						&  $0.31$, $0.93$ $\mu$m  &
		$d/2\pi$   									&  $3.25141$ kHz \\ 
		\hline
		$g$                   						&  $2$ &
		$\mu_0M_{s}$   								&  $175$ mT \\
		\hline
		$A$           		  						&  $3.7$ pJ/m &
		$E_{\text{SO}}$           		  			&  $19$ eV \\
		\hline
		$T_1$										&  $1\,\mu$s - $1$ s &
		$T_2^\ast$									&  $1$ ms - $1$ s \\
		\hline
	\end{tabular}
	\caption{\label{table:parameters}
		List of the parameters we use for our physical system, consisting of an ultrathin YIG nanostrip and a pair of NV centers placed on top of the strip.}
\end{table*}

NV center Hamiltonian describes a three-level system. The lower level is $\ket{0}$ with zero energy and upper levels are $\ket{\pm 1}$
with energies $\hbar\omega_\pm:=\hbar(D\pm\gamma_\mathrm{NV}B_0)$. In Sec.~\ref{sec:model-spinChainMagnons}, the influence of the same magnetic field $B_0$ on the spin chain has been taken into account. Consistent with our low temperature condition to develop the magnon Hamiltonian, relevant magnon states that can significantly couple to the NV center are those in the vicinity of $k=0$ mode with $\omega_0=g\mu_B B_0$. Accordingly, the relation $\omega_0 \ll \omega_+$ is satisfied with $D\gg 1$, so that we can limit the
dynamics to the manifold of $\ket{0},\ket{-1}$ and
simplify the NV center model to that of an effective two-level atom (qubit). We will consider a pair of NV center qubits, such as in two  nanodiamonds illustrated in Fig.~\ref{fig:fig1-ModelSystem}. The Hamiltonian in Eq.~(\ref{eq:Hnv}) reduces to~\cite{Rusconi2018} 
\begin{eqnarray}
\hat H_{\mathrm{NV}}=\hbar\frac{\omega_{\mathrm{NV}}}{2}\sum_{i=1,2}
\hat\sigma_i^z,
\end{eqnarray}
where $\hat{\sigma}^z_i:=|-1\rangle_i\langle -1|-|0\rangle_i\langle0|$ and 
$\omega_{\mathrm{NV}}\equiv \omega_-$. We dropped the constant terms of $\hat I\omega_-/2$ where $\hat I_i=|-1\rangle_i\langle -1|+|0\rangle_i\langle0|$ is the unit operator for each qubit. 
\section{Parameters of Physical System}\label{sec:appendix-parameters}
We present a summary of the values we used for the parameters of our physical system in Table~\ref{table:parameters}. The system consists of a YIG nanostrip subject to two external static
magnetic fields and electric field. Two diamonds hosting NV center defects are placed on top of the chain. One field is transverse to the chain and uniform. The other
field, acting on the YIG nanostrip along the chain axis but its influence on the NV centers is negligible. All the parameters are typical and accessible with the state of the art materials.

\section{Justification of the Born-Markov Approximations}\label{sec:appendix-BornMarkov}

For a typical exchange coupling coefficient $J\sim 10$ GHz and large spin $s\sim 10$, magnon subsystem has a wide bandwidth of 
$\Delta\omega=8Js\sim 10^3$ GHz. Using the dispersion relation (\ref{eq:MagnonDispersion}) and spacing between the magnon modes in the reciprocal space $\delta k = \pi/L$, we find the spacing between the
modes in the frequency space such that $\delta\omega_k=(d\omega_k/dk)\delta_k$ or $\delta\omega_k/\Delta\omega=\sin(ka)(\pi/2N)$, which allows for treating the magnon spectrum as continuous over the the bandwidth for $N\gg 1$. This justifies the Born
approximations.

\begin{figure}[b!]
	\centering
	\includegraphics[width=0.8\linewidth]{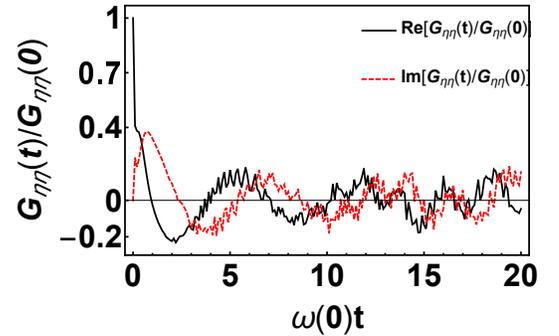}
	\caption{\label{fig:corrFun} (Color online) Real (solid black curve) and imaginary (red dashed curve) parts of the magnon bath correlation function $G_{\eta\eta}(t)$, normalized by its initial value $G_{\eta\eta}(0)$. Time $t$ is scaled with the resonance frequency $\omega(0)\equiv \omega_0\sim (2\pi) 1.4\times 10^9$ rad/s.
	}
\end{figure}

The bath correlation time can be determined by examination of the
bath correlation functions. Though we have three interaction coefficients and
a coherence function, their $k$-space widths are similar as can be seen in Figs.~\ref{fig:zetak}-\ref{fig:xik} (cf.~Fig.~\ref{fig:etak}).
We can therefore consider only one correlation function to estimate the bath correlation time, which we take
\begin{eqnarray}
G_{\eta\eta}(t):=\sum_{k=-\infty}^{\infty}|\eta_k|^2\mathrm{e}^{i\omega_k t}.
\end{eqnarray}
$G_{\eta\eta}(t)$ is plotted
in Fig.~\ref{fig:corrFun}, from which we can deduce that 
$\tau_B$ is about few nanoseconds. The correlations between the bath and the system
can build up in $\tau_B$,
but they are forgotten in longer time intervals of interest for the overall open
system dynamics. To see the relaxation time for the system $\tau_s$ to the steady state, we solve the master equation in the next section numerically. We find $\tau_s\sim$ milliseconds so that $\tau_B\ll \tau_s$. Between these two time scales, 
$\tau_B<\Delta t\ll\tau_s$, 
a coarse-grained time step $\Delta t$ can be taken and the Markov approximations
can be justified. 
\\\\

\bibliography{nvYIGmanuscript}

\end{document}